\newif\ifpreprint 
\preprinttrue

\ifpreprint 
\documentclass[a4paper]{article}
\usepackage[cm]{fullpage}
\usepackage{amsthm}
\usepackage[affil-it]{authblk}
\usepackage[table]{xcolor}
\else 
\documentclass[num-refs]{wiley-article}
\fi 

\usepackage{color}
\usepackage{ulem}

\usepackage{siunitx}
\usepackage{graphicx}
\usepackage[space]{grffile}
\usepackage{latexsym}
\usepackage{textcomp}
\usepackage{longtable}
\usepackage{tabulary}
\usepackage{booktabs,array}
\usepackage{amsfonts,amsmath,amssymb}
\usepackage{url}
\usepackage{hyperref}
\ifpreprint 
\hypersetup{colorlinks=true,bookmarksnumbered=true,citecolor=red,urlcolor=red}
\else 
\usepackage{natbib}
\hypersetup{colorlinks=false,pdfborder={0 0 0}}
\fi 
\usepackage{etoolbox}
\usepackage{enumitem}
\usepackage{amsmath,amssymb}
\usepackage{algorithm}
\usepackage{algpseudocode}
\usepackage{multirow}
\usepackage{tikz}
\usepackage{soul}
\usepackage{bm}
\usepackage{subfig}
\usepackage[utf8]{inputenc}

\newcommand\mapsfrom{\mathrel{\reflectbox{\ensuremath{\mapsto}}}}
\renewcommand{\vec}[1]{\ensuremath{\boldsymbol{#1}}}
\newcommand{\favg}[1]{\{\!\!\{#1\}\!\!\}}
\newcommand{\fjump}[1]{[\![#1]\!]}

\newcommand{\wb}{\mathbb{W}}

\newcommand{\figdir}{./figures/}
\graphicspath{\figdir}

\makeatletter
\def\blfootnotetext{\gdef\@thefnmark{}\@footnotetext}
\makeatother

\ifpreprint 
\else 
\papertype{Original Article}
\paperfield{Numerical methods and NWP}
\fi 

\title{Hybridised multigrid preconditioners for a compatible finite element dynamical core}

\ifpreprint 
\begin{document}
\else 
\fi 
\author[1]{Jack~D.~Betteridge}
\author[1]{Colin~J.~Cotter}
\author[2,6]{Thomas~H.~Gibson}
\author[3,4]{Matthew~J.~Griffith}
\author[5]{Thomas~Melvin}
\ifpreprint 
  \author[3,*]{Eike~H.~M\"{u}ller}
\else 
  \author[3]{Eike~H.~M\"{u}ller}
  \corraddress{Eike~H.~M\"{u}ller, Department of Mathematical Sciences, University of Bath, BA2 7EX Bath, United Kingdom}
  \corremail{e.mueller@bath.ac.uk}
\fi 

\affil[1]{Imperial College, London, United Kingdom}
\affil[2]{National Center for Supercomputing Applications, Urbana, IL, USA}
\affil[3]{University of Bath, UK}
\affil[4]{European Centre for Medium Range Weather Forecasts, Reading, UK}
\affil[5]{Met Office, Exeter, United Kingdom}
\affil[6]{Advanced Micro Devices, Inc., Austin, TX, United States}

\ifpreprint 
  \affil[*]{Email: \texttt{e.mueller@bath.ac.uk}}
\else 
\fi 



\ifpreprint 
  \makeatletter
  \def\blfootnote{\gdef\@thefnmark{}\@footnotetext}
  \makeatother
  \blfootnote{\textcopyright \hspace{0.5mm} Crown copyright 2020. Reproduced with the permission of the Controller of HMSO.}
\else 
  \fundinginfo{EPSRC, grant numbers EP/R511547/1, EP/R51164X/1 and EP/W522491/1}
  \blfootnotetext{\textcopyright \hspace{0.5mm} Crown copyright 2022. Reproduced with the permission of the Controller of HMSO. Published by John Wiley and Sons, Ltd.}

  \runningauthor{Betteridge et al.}
\fi 

\ifpreprint 
\else 
\begin{document}
\fi 

\maketitle

\begin{abstract}

  \ifpreprint 
    \noindent
  \else 
  \fi 
  Compatible finite element discretisations for the atmospheric equations of motion have recently attracted considerable interest. Semi-implicit timestepping methods require the repeated solution of a large saddle-point system of linear equations. Preconditioning this system is challenging since the velocity mass matrix is non-diagonal, leading to a dense Schur complement. Hybridisable discretisations overcome this issue: weakly enforcing continuity of the velocity field with Lagrange multipliers leads to a sparse system of equations, which has a similar structure to the pressure Schur complement in traditional approaches. We describe how the hybridised sparse system can be preconditioned with a non-nested two-level preconditioner. To solve the coarse system, we use the multigrid pressure solver that is employed in the approximate Schur complement method previously proposed by the some of the authors. Our approach significantly reduces the number of solver iterations. The method shows excellent performance and scales to large numbers of  cores in the Met Office next-generation climate- and weather prediction model LFRic.
  \ifpreprint 
    \\[1ex]
    \textbf{keywords}: Numerical methods and NWP, Dynamics, Atmosphere, Multigrid, Parallel Scalability, Linear Solvers, Finite Element Discretisation, Hybridisation
  \else 
  \fi 
\end{abstract}


\section{Introduction}
Over the past decade, finite element discretisations of the governing equations for atmospheric motion have become increasingly popular. A majority of the focus has been centred on high-order continuous Galerkin (CG), spectral element, and discontinuous Galerkin (DG) methods (for an overview, see \cite{fournier2004spectral, thomas2005ncar, dennis2012cam,kelly2012continuous, giraldo2013implicit, bao2015horizontally, marras2016review}). More recently, a theory of mixed-finite element discretisations, known as mimetic or compatible finite element methods \cite{cottershipton2012,staniforth2013analysis,cotterthuburn2014,mcrae2014energy,natale2016compatible,Gibson_2019}, has been developed within the context of simulating geophysical fluid dynamics. Compatible finite element elements are constructed from discrete spaces that satisfy an $L^2$ de-Rham co-homology in which differential operators surjectively map from one space to another \cite{arnold2006finite, arnold2010finite, arnold2014finite}. Discretisations arising from this framework have a long history in both numerical analysis and applications ranging from structural mechanics to porous media flows, see \cite{mixedFiniteElementsCOmpatibilityandApps, mixedfiniteelementsandApplicationsText} for a summary of contributions. The embedded property of these discrete spaces defines a finite element extension of the Arakawa C-grid staggered finite difference approaches \cite{arakawa1977computational}, while avoiding the need for discretising on structured, orthogonal meshes (such as traditional latitude-longitude meshes). The realization made by Cotter and Shipton in \cite{cottershipton2012} that compatible finite element methods respect essential geostrophic balance relations for steady-state solutions of the rotating shallow water equations, combined with other properties making them analogous to the Arakawa C-grid, has galvanized an effort to develop similar techniques for numerical weather prediction. This framework of compatible finite elements is a core area of research in the UK Met Office's effort to redesign an operational dynamical core, known as the Gung-Ho project. A key result of this project has been the development of the LFRic model \cite{melvin2019mixed,adams2019lfric}, which solves the compressible Euler equations for a perfect gas in a rotating frame. The model uses a compatible finite element discretisation on a hexahedral grid based on tensor product spaces derived from the classical Raviart-Thomas (RT) method \cite{raviartthomas1977}.
\subsection{Semi-implicit time discretisation of the compressible Euler equations}
Atmospheric flows are characterised by a separation of scales: although fast acoustic and inertia-gravity waves carry very little energy, they need to be represented in the model since they drive slower large-scale balanced flow through non-linear interactions. As a result, the timestep size in a model based on fully explicit timestepping would be prohibitively small due to the CFL condition imposed by the fast acoustic waves. To overcome this issue, the LFRic model employs a semi-implicit time discretisation which treats the fast waves implicitly in all spatial directions. This allows simulations with a significantly larger timestep size,  which can potentially reduce the overall model runtime.

The drawback of semi-implicit methods is that they require the solution of a non-linear system in every model timestep. Since the implicit system is only mildly non-linear, it can be solved efficiently with a quasi-Newton-type iteration, which in turn requires the repeated solution of a linear system. This linear solve is one of the computational bottlenecks of the LFRic model. While compatible finite element discretisations have several useful properties for geophysical flow applications, the efficient implementation of solver algorithms for the innermost linear system poses a significant challenge. This is primarily due to the fact that, unlike for standard finite difference discretisations, the linear system takes the form of an indefinite mixed problem with saddle-point structure. To deliver forecasts under strict operational time constraints, solvers must efficiently scale to many cores on distributed-memory clusters while being algorithmically robust.
\subsection{Hybridisation and related work on preconditioners}
There are a number of possible approaches for handling the linear saddle-point system arising from compatible finite element discretisations. This includes H(div)-multigrid \cite{arnoldfalkwinther2000}, requiring complex overlapping Schwarz smoothers; auxiliary space multigrid \cite{hiptmairxu2007}; and global Schur-complement preconditioners, requiring inverting both a velocity mass matrix and an elliptic Schur-complement operator. So far, a pragmatic solution has been employed in LFRic: the saddle-point system is preconditioned with an approximate Schur-complement method which is obtained by lumping the velocity mass matrix \cite{Maynard2020,mitchell2016high}. The resulting elliptic Schur-complement system in the piecewise constant pressure space is solved with a bespoke multigrid algorithm that exploits the tensor-product structure of the function spaces, following \cite{Borm2001}. The drawback of this approach is that it requires the repeated, expensive application of the saddle-point operator for all four prognostic variables in a preconditioned Krylov-iteration.

To address this issue, we focus on a technique known as hybridisation \cite{arnoldbrezzi1985, cockburnandjay2004} in this paper. Hybridisation utilises element-wise static condensation to obtain a reduced problem on the mesh interfaces. Once solved, recovery of the prognostic variables is achieved by solving purely local linear systems. Although the reduced, sparse system is smaller than the original saddle-point system for the prognostic variables, it has to be solved efficiently. Non-nested algorithms \cite{bramble1991analysis} have been proposed for the solution of linear systems that arise from the hybridisation of mimetic finite element discretisations. The authors of \cite{Gopalakrishnan2009} describe a non-nested multigrid method for the solution of the hybridisable mixed formulation of a second order elliptic boundary value problem, while \cite{cockburn2014multigrid} use a similar approach for the DG discetisation of the same equation. Both papers employ a conforming coarse level space based on piecewise linear elements; this space allows the construction of a standard multigrid method with $h$-coarsening. The hybridised method has also been employed to solve the linear system that arises in an IMEX-DG discretisation of the shallow water equations in \cite{kang2020imex}. While the authors of \cite{kang2020imex} use a direct solver for the resulting elliptic system on the mesh interfaces, a non-nested multigrid algorithm similar to \cite{cockburn2014multigrid} is used for the IMEX-DG discretisation of the shallow water equations in \cite{betteridge2021multigrid}.
\subsection{Main contributions}
In this paper, we describe how a hybridisable discretisation of the compressible Euler equations in LFRic allows the efficient solution of the saddle-point system for the prognostic variables if it is combined with a non-nested multigrid preconditioner. Our work is motivated by two scientific questions:
\begin{enumerate}
  \item Can hybridisation be used to solve the saddle-point system in LFRic more efficiently?
  \item What performance gains can be achieved by using bespoke non-nested multigrid preconditioners?
\end{enumerate}
The new contributions of our work are:
\begin{enumerate}[label=\roman*.]
  \item We decribe how the exact elimination of density and potential temperature allows the construction of a reduced saddle-point system for Exner pressure and momentum; solving this system with the approximate Schur-complement approach with a pressure multigrid preconditioner improves on the results in \cite{Maynard2020} and forms the basis for our hybridisable discretisation.
  \item We derive a hybridisable discretisation of the compressible Euler equations and use static condensation to construct an elliptic system for the unknowns on the mesh skeleton.
  \item We introduce a non-nested multigrid algorithm for the solution of this elliptic system; since the unknowns on the mesh skeleton approximate pressure, we use the pressure multigrid algorithm from \cite{Maynard2020} to solve the piecewise constant pressure equation on the coarse level of the multigrid hierarchy.
  \item We compare the performance of different solver setups to demonstrate the impact of hybridisation and the use of a multigrid preconditioner, thus providing empirical evidence to answer our two central scientific questions above.
  \item We demonstrate the excellent parallel scalability of our implementation on up to 13824 cores.
\end{enumerate}
The choice of a piecewise constant coarse level space in the non-nested multigrid preconditioner is mainly motivated by practical considerations, since it allows the re-use of the efficient pressure multigrid algorithm in \cite{Maynard2020}. Although a rigorous theoretical justification of this approach is beyond the scope of this paper, our numerical results show that it works very well in practice. While for the present, lowest order discretisation employed in LFRic, our hybridised solver is not able to beat the optimised approximate Schur-complement approach in \cite{Maynard2020}, we will argue in Section \ref{sec:conclusion} that this picture is likely to change for higher-order discretisations. This is due to the dramatic increase in condition number and problem size for the resulting mixed operators. In situations like this, it has been show that static condensation approachs can provide significant benefit \cite{Pardo_2015}. We therefore believe that the hybridised solver introduced in this paper has significant potential to accelerate semi-implicit compatible finite element models for atmospheric fluid dynamics.
\paragraph{Structure.}
The rest of this paper is organised as follows: in Section \ref{sec:methods} we describe the mixed finite element discretisation of the compressible Euler equations that is used in LFRic and show how potential temperature and density can be eliminated to obtain a smaller linear system for Exner pressure and velocity only. We also explain how hybridisation allows the construction of a sparse Schur complement system. After reviewing the approximate Schur-complement solver and pressure multigrid algorithm from \cite{Maynard2020}, a new non-nested multigrid algorithm for solving this system is introduced in this section. Following a description of our implementation and the model setup, we compare the performance of different solver algorithms for a baroclinic test case in Section \ref{sec:results}, which also contains results from massively parallel scaling runs. We conclude and outline avenues for further work in Section \ref{sec:conclusion}. Further technical details on the LFRic implementation and parameter tuning can be found in the appendices.
\section{Methods}\label{sec:methods}
\subsection{Model equations} \label{sec:model_eqns}
To obtain an update for the non-linear outer iteration in a semi-implicit timestepping scheme we need to compute increments in Exner pressure $\Pi'$, potential temperature $\theta'$, density $\rho'$ and velocity $\vec{u}'$. This can be achieved by linearising the compressible Euler equations around a suitable reference state and solving the following four equations:
\begin{subequations}
  \begin{eqnarray}
    \left[\vec{u}'+\tau_u\Delta t\left(\mu\widehat{\vec{z}}(\widehat{\vec{z}}\cdot\vec{u}') + 2\vec{\Omega}\times\vec{u}'\right)\right] + \tau_u \Delta t c_p \left(\theta'\nabla\Pi^* + \theta^*\nabla\Pi'\right) &=& \vec{r}_u,\label{eqn:gh-u}\\
    \rho' + \tau_\rho\Delta t\nabla\cdot\left(\rho^*\vec{u}'\right) &=& r_\rho,\label{eqn:gh-rho}\\
    \theta' + \tau_\theta\Delta t\vec{u}'\cdot\nabla\theta^* &=& r_\theta,\label{eqn:gh-theta}\\
    \frac{1-\kappa}{\kappa}\frac{\Pi'}{\Pi^*} - \frac{\rho'}{\rho^*} - \frac{\theta'}{\theta^*} &=& r_\Pi.\label{eqn:gh-pi}
  \end{eqnarray}
\end{subequations}
Here $\Delta t$ is the timestep size, $\tau_{u},\tau_{\rho},\tau_{\theta}$ are relaxation parameters, the unit normal in the vertical direction is denoted by $\widehat{\vec{z}}$, and we have $\mu=1+\tau_u\tau_\theta \Delta t^2 N^2$ with the Brunt-V\"{a}is\"{a}l\"{a} frequency $N^2$. The right-hand sides $\vec{r}_u$, $r_\rho$, $r_\theta$, and $r_\Pi$ are residuals obtained from the linearisation process. In LFRic the reference states $\theta^*$, $\rho^*$ and $\Pi^*$ are the fields at the previous timestep. The expressions on the right-hand side depend on the current non-linear iterate, but their exact form is not relevant for the following derivation. The equations have to be solved in a thin spherical shell $\Omega$, with suitable boundary conditions at the top and bottom of the atmosphere. For a further discussion of the equations we refer the reader to \cite{Wood2014} and \cite[Section 3.3]{Maynard2020}. It should be noted that in \cite{Maynard2020}, Eqs. \eqref{eqn:gh-u} - \eqref{eqn:gh-pi} are discretised to obtain a $4\times4$ block matrix system, which is solved with a suitable preconditioned Krylov subspace method. Since each application of the $4\times4$ block matrix is expensive, we pursue a different approach here: by eliminating $\theta'$ and $\rho'$ first, we obtain a smaller $2\times2$ block system for velocity $\vec{u}'$ and Exner pressure $\Pi'$ only. This system is then solved with a preconditioned Krylov subspace method or the hybridised approach described in Section \ref{sec:hybridisation}, followed by the reconstruction of $\theta'$ and $\rho'$ from $\vec{u}'$ and $\Pi'$. Since the elimination and recovery of $\theta'$ and $\rho'$ is only necessary once per linear solve, this increases efficiency.
\paragraph{Analytic elimination of potential temperature.}
Eq. \eqref{eqn:gh-theta} can be used to eliminate $\theta'$ from Eqs. \eqref{eqn:gh-u}, \eqref{eqn:gh-rho} and \eqref{eqn:gh-pi}, leading to the following system of three equations:
\begin{subequations}
  \begin{eqnarray}
    \left[\vec{u}'+\tau_u\Delta t\left(\mu\widehat{\vec{z}}(\widehat{\vec{z}}\cdot\vec{u}') + 2\vec{\Omega}\times\vec{u}'\right)
      -\tau_u \tau_\theta \Delta t^2 c_p \left(\vec{u}'\cdot\nabla \theta^*\right)\nabla\Pi^*\right] + \tau_u \Delta t c_p  \theta^*\nabla\Pi' &=& \overline{\vec{r}}_u\equiv \vec{r}_u-\tau_u\Delta t c_p \nabla \Pi^* r_\theta,\label{eqn:gh-u_elim}\\
    \rho' + \tau_\rho\Delta t\nabla\cdot\left(\rho^*\vec{u}'\right) &=& r_\rho,\label{eqn:gh-rho_elim}\\
    \frac{1-\kappa}{\kappa}\frac{\Pi'}{\Pi^*} - \frac{\rho'}{\rho^*} + \tau_\theta\Delta t\frac{\vec{u}'\cdot\nabla \theta^*}{\theta^*} &=& \overline{r}_\Pi\equiv r_\Pi + \frac{r_\theta}{\theta^*}\label{eqn:gh-pi_elim}.
  \end{eqnarray}
\end{subequations}

\subsection{Mixed finite element discretisation}\label{sec:mixed_finite_element_discretisation}
To discretise Eqs. \eqref{eqn:gh-u_elim} - \eqref{eqn:gh-pi_elim}, we proceed as in \cite[Section 3.2]{Maynard2020} and introduce finite element spaces on a hexahedral grid $\Omega_h$ covering the domain $\Omega$: $\mathbb{W}_2$ is the Raviart-Thomas \cite{Raviart1977} space $RT_p$ and $\mathbb{W}_3=Q_p^{\text{DG}}$ is the scalar discontinuous Galerkin space, where in both cases $p$ is the order of the discretisation. The velocity space $\mathbb{W}_2=\mathbb{W}_2^\text{h}\oplus \mathbb{W}_2^\text{z}$ can be written as the direct sum of a space $\mathbb{W}_2^h$, which only contains vectors pointing in the horizontal, and a space $\mathbb{W}_2^\text{z}$ with purely vertical fields (see \cite[Figure 1]{Maynard2020}). Recall further that the space $\mathbb{W}_2^\text{z}$ is continuous in the vertical direction, but discontinuous in the horizontal direction, while $\mathbb{W}_2^\text{h}$ is horizontally continuous and vertically discontinuous. To represent the reference profile for potential temperature, a third space $\mathbb{W}_\theta$ is introduced. $\mathbb{W}_\theta$ is the scalar-valued equivalent of $\mathbb{W}_2^\text{z}$ and has the same continuity, which is analogous to using a Charney-Phillips staggered vertical discretisation. A further discussion of the function spaces can be found in \cite{melvin2019mixed}.

We now assume that $\vec{u}'\in\mathbb{W}_2$, $\Pi',\rho',\Pi^*,\rho^*\in\mathbb{W}_3$ and $\theta^*\in\mathbb{W}_\theta$. Multiplying Eqs. \eqref{eqn:gh-u_elim} - \eqref{eqn:gh-pi_elim} by suitable test functions $\vec{w}\in\mathbb{W}_2$ and $\phi,\psi \in\mathbb{W}_3$ and integrating over $\Omega_h$ leads to the following weak form of the problem:

\begin{equation}
  \begin{aligned}
    \breve{m}_2(\vec{w},\vec{u}') - q_{22}(\vec{w},\vec{u}') + g(\vec{w},\Pi')
                                                                      & = r_u(\vec{w}),                                                                                    \\
    d(\phi,\vec{u}') + m_3(\phi,\rho')                                & = r_\rho(\phi),\qquad\text{for all $\vec{w}\in \mathbb{W}_2$ and all $\phi,\psi\in \mathbb{W}_3$.} \\
    q_{32}(\psi,\vec{u}') - m_3^\rho(\psi,\rho') + m_3^\Pi(\psi,\Pi') & = r_\Pi(\psi),
  \end{aligned}
  \label{eqn:weak_form}
\end{equation}
Writing $\langle\cdot\rangle_{\Omega_h}$ for integrals over the mesh $\Omega_h$ and $\langle\cdot \rangle_{\mathcal{E}_h}$ for integrals over the mesh skeleton $\mathcal{E}_h$, i.e.
\begin{xalignat*}{2}
  \langle f \rangle_{\Omega_h} &=
  \sum_{K \in \Omega_h} \int_K f \;dx,&
  \langle g \rangle_{\mathcal{E}_h} &=
  \sum_{\partial K \in \mathcal{E}_h} \int_{\partial K} g \; dS,
\end{xalignat*}
the eight bilinear forms on the left-hand side of Eq. \eqref{eqn:weak_form} are
\begin{equation}
  \begin{aligned}
    \breve{m}_2(\vec{w},\vec{u}') & =
    \left\langle \vec{w}\cdot\left(\vec{u}'+\tau_u \Delta t \left(\mu\widehat{\vec{z}}(\widehat{\vec{z}}\cdot\vec{u}')+2\vec{\Omega}\times\vec{u}'\right)\right)\right\rangle_{\Omega_h},                                    \\
    q_{22}(\vec{w},\vec{u}')      & = \tau_u\tau_\theta \Delta t^2 \left\langle c_p (\widehat{\vec{z}}\cdot\vec{w})\partial_z \widehat{\Pi}^* (\widehat{\vec{z}}\cdot \vec{u}')\partial_z \theta^* \right\rangle_{\Omega_h}, \\
    g(\vec{w},\Pi')               & = \tau_u\Delta t c_p \left(\left\langle\fjump{\theta^*\vec{w}}\favg{\Pi'}\right\rangle_{\mathcal{E}_h}-\left\langle\nabla\cdot(\theta^*\vec{w})\Pi'\right\rangle_{\Omega_h}\right),      \\
    d(\phi,\vec{u}')              & = \tau_\rho \Delta t\left\langle \phi \nabla\cdot(\rho^*\vec{u}')\right\rangle_{\Omega_h},                                                                                               \\
    m_3(\phi,\rho')               & = \left\langle \phi \rho'\right\rangle_{\Omega_h},                                                                                                                                       \\
    q_{32}(\psi,\vec{u}')         & = \tau_\theta \Delta t\left\langle\psi \frac{\partial_z \theta^*}{\theta^*}(\widehat{\vec{z}}\cdot \vec{u}')\right\rangle_{\Omega_h},                                                    \\
    m_3^\rho(\psi,\rho')          & =  \left\langle\frac{\psi\rho'}{\rho^*}\right\rangle_{\Omega_h},                                                                                                                         \\
    m_3^\Pi(\psi,\Pi')            & = \frac{1-\kappa}{\kappa} \left\langle\frac{\psi\Pi'}{\Pi^*}\right\rangle_{\Omega_h}.
  \end{aligned}\label{eqn:bilinear_forms}
\end{equation}
Since it is important for the discussion of hybridisation in Section \ref{sec:hybridisation}, we explain the derivation of the weak form $g(\cdot,\cdot)$ in more detail. For this, consider the term $\theta^*\nabla \Pi'$ in Eq. \eqref{eqn:gh-u}. The function space $\mathbb{W}_3$ is discontinuous and the derivative of $\Pi'$ can not be evaluated directly. Instead, we integrate the term by parts after it has been multiplied  by a test function $\vec{w} \in \mathbb{W}_2$. Considering a single cell $K$ of the mesh, this leads to two integrals:
\begin{equation}
  \int_K \theta^*\vec{w}\cdot\nabla \Pi' \;dx
  = -\int_K \nabla\cdot \left(\theta^* \vec{w}\right) \Pi' \;dx
  + \int_{\partial K} \left(\theta^*\vec{w}\cdot \vec{n}\right)\Pi'|_{\partial K}\;dS,
  \label{eqn:single_cell_integral_non-hybridised}
\end{equation}
where $\vec{n}$ is the unit outward normal vector at each point of the cell surface $\partial K$. In the surface integral, we need to decide how to evaluate the field $\Pi'|_{\partial K}$, which is not well-defined since $\mathbb{W}_3$ is discontinuous. The natural choice is to take the average $\favg{\Pi'} := \frac{1}{2}\left(\Pi'_++\Pi'_-\right)$ of the fields in the two cells $K^+$, $K^-$ that are adjacent to a particular facet. Summing the right-hand side of Eq. \eqref{eqn:single_cell_integral_non-hybridised} over all cells of the mesh then leads to the expression in $g(\cdot,\cdot)$, namely
\begin{equation}
  -\left\langle\nabla\cdot\left(\theta^*\vec{w}\right)\Pi'\right\rangle_{\Omega_h}
  + \left\langle\fjump{\theta^*\vec{w}}\favg{\Pi'}\right\rangle_{\mathcal{E}_h}.
  \label{eqn:g_with_pi_avg}
\end{equation}
In this expression $\fjump{\theta^*\vec{w}} \equiv \theta^*_+ (\vec{w}_+\cdot \vec{n}_+) + \theta^*_- (\vec{w}_-\cdot \vec{n}_-)$ and the subscripts ``$+$'' and ``$-$'' are again used to identify quantities defined on the two cells that are adjacent to a particular facet.  Note that although here $\vec{w}_+\cdot \vec{n}_+=-\vec{w}_-\cdot\vec{n}_-$ as $\vec{n}_+=-\vec{n}_-$ and the normal components of the velocity field are continuous across facets, this is no longer true in our hybridisable formulation (see Section \ref{sec:hybridisation}).

To obtain the bilinear forms $q_{22}(\cdot,\cdot)$ and $q_{32}(\cdot,\cdot)$, the following approximations have been made: when evaluating $\nabla \Pi^*$ in Eq. \eqref{eqn:gh-u_elim}, the reference profile is projected into the function space $\mathbb{W}_\theta$ to obtain $\widehat{\Pi}^*\in\mathbb{W}_\theta$. Since $\mathbb{W}_\theta$ is only continuous in the vertical direction, the horizontal components of the derivatives $\nabla \theta^*$ and $\nabla \Pi^*$ in Eqs. \eqref{eqn:gh-u_elim} and \eqref{eqn:gh-pi_elim} are dropped to obtain $\widehat{\vec{z}} \partial_z \theta^*$ and $\widehat{\vec{z}} \partial_z \widehat{\Pi}^*$. This is a sensible approximation because the reference profiles vary predominantly in the vertical direction.

The linear forms on the right-hand side of Eq. \eqref{eqn:weak_form} are given by $r_u(\vec{w})=\left\langle\vec{w}\cdot \overline{\vec{r}}_u\right\rangle_{\Omega_h}$, $r_\rho(\phi)=\left\langle\phi r_\rho \right\rangle_{\Omega_h}$ and $r_\Pi(\psi)=\left\langle\psi \overline{r}_\Pi\right\rangle_{\Omega_h}$.

In general, the integrals in Eq. \eqref{eqn:bilinear_forms} are approximated by $n$-point Gaussian quadrature rules (with $n=p+3$ for an element of order $p$), in each coordinate direction for the volume integrals $\langle\cdot \rangle_{\Omega_h}$ and in two dimensions restricted to each facet for the facet integrals $\langle\cdot\rangle_{\mathcal{E}_h}$. This value of $n$ is chosen as it is generally sufficient to integrate the terms in Eq. \eqref{eqn:bilinear_forms} exactly on an affine mesh. However, since it desirable for the equation of state $r_\Pi$ in Eq. \eqref{eqn:gh-pi_elim} to hold exactly, the bilinear forms $q_{32}$, $m_3^\rho$ and $m_3^\Pi$ in Eq. \eqref{eqn:bilinear_forms} are instead evaluated at the nodal points of the $\mathbb{W}_3$ space, such that the diagnostic relationship $r_\Pi$ between $\rho,\,\theta,\,\Pi$ holds exactly at these nodal points. For $p=0$ elements this can be simply enforced by using a one-point quadrature rule with the quadrature point located in the cell centre. Since for $p=0$ the test functions $\psi\in\mathbb{W}_3$ are constant, this is equivalent to enforcing Eq. \eqref{eqn:gh-pi_elim} in strong form at the quadrature points.
\paragraph{Matrix formulation.}
Choosing suitable basis functions $\vec{v}_j(\vec{\chi})\in\mathbb{W}_2$, $\sigma_j(\vec{\chi})\in\mathbb{W}_3$ that depend on the spatial coordinate $\vec{\chi}\in\Omega_h$, the fields $\vec{u}'$, $\rho'$ and $\Pi'$ can be expanded as
\begin{xalignat}{3}
  \vec{u}'(\vec{\chi}) &= \sum_j \widetilde{u}'_j\vec{v}_j(\vec{\chi})\in\mathbb{W}_2, &
  \rho'(\vec{\chi}) &= \sum_j \widetilde{\rho}'_j\sigma_j(\vec{\chi})\in\mathbb{W}_3, &
  \Pi'(\vec{\chi}) &= \sum_j \widetilde{\Pi}'_j\sigma_j(\vec{\chi})\in\mathbb{W}_3,
  \label{eqn:basis_functions}
\end{xalignat}
where for each quantity $a'$ the corresponding dof (degrees of freedom)-vector is given by $\widetilde{a}' = [a_1, a_2, \dots]$. This allows expressing the weak form in Eq. \eqref{eqn:weak_form} as a $3\times 3$ matrix equation for the dof-vectors $\widetilde{u}'$, $\widetilde{\rho}'$ and $\widetilde{\Pi}'$:
\begin{equation}
  \begin{pmatrix}
    \breve{M}_2 - Q_{22} & 0          & G       \\
    D                    & M_3        & 0       \\
    Q_{32}               & - M_3^\rho & M_3^\Pi
  \end{pmatrix}
  \begin{pmatrix}
    \widetilde{u}' \\\widetilde{\rho}'\\\widetilde{\Pi}'
  \end{pmatrix}
  =
  \begin{pmatrix}
    \mathcal{R}_{u} \\\mathcal{R}_\rho\\\mathcal{R}_{\Pi}
  \end{pmatrix}.\label{eqn:3x3matrix}
\end{equation}
The entries of the matrices $\breve{M}_2$, $Q_{22}$, $G$, $D$, $M_3$, $Q_{32}$, $M_3^\rho$ and $M_3^\Pi$ are obtained by evaluating the bilinear forms in Eq. \eqref{eqn:bilinear_forms} for the basis functions, for example
$(\breve{M}_{2})_{ij}=\breve{m}_{2}(\vec{v}_i,\vec{v}_j)$ and
$G_{ij}=g(\vec{v}_i,\sigma_j)$. Similarly, the components of the
residual (dual) vectors $\mathcal{R}_{u}$, $\mathcal{R}_\rho$ and $\mathcal{R}_{\Pi}$ on the right-hand side of Eq. \eqref{eqn:3x3matrix} are obtained by evaluating the linear forms $r_u$, $r_\rho$ and $r_\Pi$ for the basis functions, for example $(\mathcal{R}_u)_i = r_u(\vec{v}_i)$.
\paragraph{Elimination of density.}
Finally, the density $\widetilde{\rho}'$ is eliminated algebraically from Eq. \eqref{eqn:3x3matrix} by using $\widetilde{\rho'}=M_3^{-1}(\mathcal{R}_\rho-D\widetilde{u'})$ to obtain a $2\times 2$ matrix system for velocity $\widetilde{u}'$ and Exner pressure $\widetilde{\Pi}'$ only:
\begin{equation}
  \begin{pmatrix}
    \breve{M}_2 - Q_{22}         & G       \\
    Q_{32} + M_3^\rho M_3^{-1} D & M_3^\Pi
  \end{pmatrix}
  \begin{pmatrix}
    \widetilde{u}' \\\widetilde{\Pi}'
  \end{pmatrix}
  =
  \begin{pmatrix}
    \mathcal{R}_{u} \\\overline{\mathcal{R}}_{\Pi}
  \end{pmatrix}
  \equiv
  \begin{pmatrix}
    \mathcal{R}_{u} \\\mathcal{R}_{\Pi}+M_3^\rho M_3^{-1}\mathcal{R}_\rho
  \end{pmatrix}\label{eqn:2x2system}
\end{equation}
In the next section, we review a suitable preconditioned Krylov subspace method for solving Eq. \eqref{eqn:2x2system} (namely the approximate Schur complement pressure multigrid method in \cite{Maynard2020}) and then introduce an alternative solution method based on hybridisation in Section \ref{sec:hybridisation}.
\subsection{Pressure multigrid for approximate Schur complement solver}\label{sec:approximate_schur_multigrid}
The traditional approach for solving Eq. \eqref{eqn:2x2system}, which is pursued in the ENDGame dynamical core \cite{Wood2014}, is to eliminate the velocity $\widetilde{u}'$ and then solve the resulting elliptic Schur complement equation for the Exner pressure increment $\widetilde{\Pi}'$. For the finite element discretisation considered here, however, the matrix $\breve{M}_2-Q_{22}$ is not diagonal. This makes this approach unfeasible since the resulting Schur complement would be dense because it contains the inverse of $\breve{M}_2-Q_{22}$. As in \cite{Maynard2020}, we address this issue by constructing an \textit{approximate} Schur complement which is sparse, and then using this to precondition a Krylov-subspace iteration over the mixed system in Eq. \eqref{eqn:2x2system}. To construct the approximate Schur complement, we replace $\breve{M}_2-Q_{22}$ by a suitable lumped, diagonal approximation
$\mathring{M}_2$, which results in the following $2\times 2$ system:
\begin{equation}
  \begin{pmatrix}
    \mathring{M}_2               & G       \\
    Q_{32} + M_3^\rho M_3^{-1} D & M_3^\Pi
  \end{pmatrix}
  \begin{pmatrix}
    \widetilde{u}' \\\widetilde{\Pi}'
  \end{pmatrix}
  =
  \begin{pmatrix}
    \mathcal{R}_{u} \\\overline{\mathcal{R}}_{\Pi}
  \end{pmatrix}\label{eqn:2x2system_approx}.
\end{equation}
Eq. \eqref{eqn:2x2system_approx} has a sparse Schur complement and can be solved as follows:
\begin{description}
  \item[Step 1:] Compute the modified right-hand side
    \begin{equation}
      \mathcal{B} = \overline{\mathcal{R}}_{\Pi} - \left(Q_{32} + M_3^\rho M_3^{-1} D\right)\mathring{M}_2^{-1}\mathcal{R}_u
      \label{eqn:pressure_rhs}
    \end{equation}
  \item[Step 2:] Solve the sparse elliptic system
    \begin{equation}
      H\widetilde{\Pi}' = \mathcal{B}\qquad\text{with}\quad
      H = -\left(Q_{32} + M_3^\rho M_3^{-1} D\right)\mathring{M}_2^{-1}G + M_3^\Pi
      \label{eqn:elliptic_pressure_system}
    \end{equation}
  \item[Step 3:] Recover the velocity using
    \begin{equation}
      \widetilde{u}' = \mathring{M}_2^{-1}(\mathcal{R}_u-G\widetilde{\Pi}')
      \label{eqn:velocity_recovery}
    \end{equation}
\end{description}
To (approximately) solve the elliptic  system in Eq. \eqref{eqn:elliptic_pressure_system} we use a tensor-product multigrid \cite{Borm2001} approach. Due to the high aspect ratio of the grid ($\Delta z\ll \Delta x$), the Helmholtz operator is dominated by couplings within each vertical column. A suitable operator $\widehat{H}_z\approx H$, which captures these columnwise couplings, is constructed as in \cite[Section 3.6]{Maynard2020} and used as a block-Jacobi smoother in the multigrid iteration:
\begin{equation}
  \widetilde{\Pi}' \mapsfrom \widetilde{\Pi}'+\omega \widehat{H}_z^{-1}\left(\mathcal{B}-H\widetilde{\Pi}'\right),
  \label{eqn:pressure_block_jacobi}
\end{equation}
where $\omega$ is a relaxation factor. Since the matrix $\widehat{H}_z$ is block-diagonal, it can be inverted independently in each vertical column with the Thomas algorithm \cite{Press2007}. We refer the reader to \cite{Maynard2020} for further details, such as the construction of suitable prolongation and restriction operators.
\subsection{Hybridisable formulation}\label{sec:hybridisation}
\begin{figure}
  \begin{center}
    \includegraphics[width=0.5\linewidth]{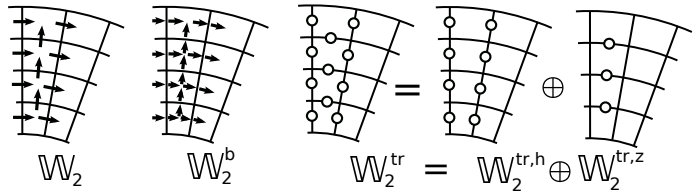}
    \caption{Function space degrees of freedom (DOFs) used for hybridisable formulation.
    The original velocity function space $\mathbb{W}_2$ (left) is replaced by the broken function space $\mathbb{W}_2^b$ (centre). The trace function space $\mathbb{W}_2^{\text{tr}}$ (right) can be decomposed into $\mathbb{W}_2^{\text{tr,h}}$ and $\mathbb{W}_2^{\text{tr,z}}$; compare to the decomposition of $\mathbb{W}_2$ in \cite[Figure 1]{Maynard2020}. Arrows indicate DOFs evaluating normal components of vectors, and points indicate evaluation of scalars.}
    \label{fig:function_spaces}
  \end{center}
\end{figure}

The main drawback of the approximate Schur complement method in Section \ref{sec:approximate_schur_multigrid} is that it still requires an iterative scheme to be applied to the mixed operator in Eq. \eqref{eqn:2x2system} which is expensive: at lowest order, on average four unknowns are stored per grid cell to represent the pressure and velocity unknowns (one pressure unknown per cell plus one shared velocity unknown for each of the six facets). The mixed operator is stored and applied in the form of local stiffness matrices and since there are seven unknowns per grid cell (one pressure unknown and six velocity unknowns), each of these matrices has $7^2 = 49$ unknowns, which makes the operator application expensive. The hybridisable approach presented here seeks to address these issues. Recall that the continuity of the velocity space prevents the construction of a sparse Schur complement. We therefore introduce a broken function space $\mathbb{W}_2^b$ for the velocity field such that in each grid cell $\mathbb{W}_2^b$ is constructed from the same local basis elements as  $\mathbb{W}_2$, but --- in contrast to $\mathbb{W}_2$ --- does not necessarily have a continuous normal component across neighbouring cells, see Figure \ref{fig:function_spaces}.
We expand the broken velocity field $\vec{u}'^{b}\in\mathbb{W}_2^b$ as in Eq. \eqref{eqn:basis_functions}
\begin{equation}
  \vec{u}'^b(\vec{\chi}) = \sum_j \widetilde{u}'^b_j\vec{v}^b_j(\vec{\chi})\in\mathbb{W}_2^b,
  \label{eqn:broken_basis_functions}
\end{equation}
where now each basis function $\vec{v}^b_j(\vec{\chi})$ is non-zero only in a single grid cell $K$. More specifically, each $\vec{v}^b_j(\vec{\chi})$ is obtained by multiplying a Raviart Thomas basis function with the indicator function $\chi_K(\vec{\chi})$ which satifies $\chi_K(\vec{\chi}) = 1$ for $\vec{\chi}\in K$ and $\chi_K(\vec{\chi})=0$ otherwise. To weakly enforce the continuity of the velocity fields, Lagrange multipliers are introduced. This leads to an additional equation, which, if solved exactly, ensures that the velocity field is projected into the subspace $\mathbb{W}_2\subset \mathbb{W}_2^b$ such that $\mathbb{W}_2$ has continuous normal components across facets.

Replacing $\mathbb{W}_2$ by $\mathbb{W}_2^b$ affects four of the bilinear forms in Eq. \eqref{eqn:bilinear_forms}. The forms $\breve{m}_2(\cdot,\cdot)$, $d(\cdot,\cdot)$ and $q_{32}(\cdot,\cdot)$ are volume integrals, and evaluating them for the basis functions $\vec{v}_j^b$ leads to three matrices $\breve{M}_{2}^b$, $D^b$ and $Q_{32}^b$ with, for example $D^b_{ij} = d(\sigma_i,\vec{v}_j^b)$. It is important to note that with this choice of basis all three matrices are block-diagonal, coupling only unknowns in a single grid cell. This also implies that $\breve{M}_{2}^b$, $D^b$ and $Q_{32}^b$ can be inverted cell-wise. The bilinear form $g(\cdot,\cdot)$ requires more thought. Instead of constructing it by using the average $\favg{\Pi'}$ in the facet integral in Eq. \eqref{eqn:g_with_pi_avg}, we introduce a new field $\lambda\in\mathbb{W}_2^{\text{tr}}$ which is an approximation of the Exner pressure increment in the continuous equations. The trace space $\mathbb{W}_2^{\text{tr}}$ is a scalar valued function space with support on the skeleton $\mathcal{E}_h$. It is piecewise continuous, with each function being represented by a polynomial of degree $p$ on an individual facet. Consequently, the facet integral in Eq. \eqref{eqn:g_with_pi_avg} becomes $\left\langle\right\fjump{\theta^*\vec{w}}\lambda\rangle_{\mathcal{E}_h}$ and the bilinear form $g(\vec{w},\Pi')$ in Eq. \eqref{eqn:bilinear_forms} is replaced by
\begin{equation}
  g^b(\vec{w}^b,\Pi') + k(\vec{w}^b,\lambda) \qquad \text{for $\vec{w}^b\in\mathbb{W}_2^b$, $\Pi\in\mathbb{W}_3$ and $\lambda\in\mathbb{W}_2^{\text{tr}}$}
\end{equation}
with
\begin{equation}
  \begin{aligned}
    g^b(\vec{w}^b,\Pi')  & = -\tau_u\Delta t c_p \left\langle\nabla\cdot(\theta^*\vec{w}^b)\Pi'\right\rangle_{\Omega_h},  \\
    k(\vec{w}^b,\lambda) & = \tau_u\Delta t c_p\left\langle\fjump{\theta^*\vec{w}^b}\lambda\right\rangle_{\mathcal{E}_h}. \\
  \end{aligned}
\end{equation}
Finally, continuity of the normal components of the velocity field
is enforced weakly through the equation
\begin{equation}
  c(\vec{u}'^b,\nu) = \left\langle \fjump{\vec{u}'^b}\nu\right\rangle_{\mathcal{E}_h} = 0 \qquad\text{for all $\nu\in\mathbb{W}_2^{\text{tr}}$}.
\end{equation}
This is the crucial step that makes the hybridisable method consistent with the mixed method that it is based on, since it ensures that the resulting velocity field is H(div) conforming.

As above, matrix representations $G^b$, $K$ and $C$ of the bilinear forms $g^b(\cdot,\cdot)$, $k(\cdot,\cdot)$ and $c(\cdot,\cdot)$ are obtained by evaluating them for the basis functions. This leads to the following matrix equation
\begin{equation}
  \begin{pmatrix}
    \breve{M}^b_u - Q_{22}^b      & G^b     & K \\
    Q_{32}^b+M_3^\rho M_3^{-1}D^b & M_3^\Pi & 0 \\
    C^T                           & 0       & 0
  \end{pmatrix}
  \begin{pmatrix}
    \widetilde{u}'^b \\ \widetilde{\Pi}' \\ \widetilde{\lambda}
  \end{pmatrix}
  =
  \begin{pmatrix}
    \mathcal{R}^b_u \\ \overline{\mathcal{R}}_\Pi\\ 0
  \end{pmatrix}\label{eqn:gh-fem-hybrid}.
\end{equation}
Since $\mathbb{W}_2\subset \mathbb{W}_2^b$, the vector $\mathcal{R}_u^b$ on the right-hand side is readily obtained from $\mathcal{R}_u$ with a simple basis transformation. The $3\times 3$ system in Eq. \eqref{eqn:gh-fem-hybrid} is larger than the original $2\times 2$ system in Eq. \eqref{eqn:2x2system}, both due to the fact that $\wb_2^b$ has more unknowns than $\mathbb{W}_2$ and since the Lagrange multipliers $\lambda$ need to be solved for as well. However, the $3\times 3$ matrix in Eq. \eqref{eqn:gh-fem-hybrid} has the crucial advantage that its upper left $2\times 2$ submatrix has a block-diagonal structure, with each block only coupling the unknowns in a single grid cell. This implies that it can be inverted very efficiently by inverting small, cell-local matrices. Furthermore, eliminating velocity and pressure from Eq. \eqref{eqn:gh-fem-hybrid} results in a (block-)sparse Schur complement for $\widetilde{\lambda}$, and the velocity $\widetilde{u}'$ and Exner pressure $\widetilde{\Pi}'$ can be recovered with purely local operations if the dof-vector $\widetilde{\lambda}$ is known.
To make this more explicit, we rewrite Eq. \eqref{eqn:gh-fem-hybrid} as
\begin{equation}
  \begin{pmatrix}
    \mathbf{A}   & \mathbf{K} \\
    \mathbf{C}^T & 0
  \end{pmatrix}
  \begin{pmatrix}
    \widetilde{\mathbf{X}}' \\ \widetilde{\lambda}
  \end{pmatrix}
  =
  \begin{pmatrix}
    \vec{\mathcal{R}}_X \\ 0
  \end{pmatrix}\label{eqn:gh-fem-hybrid-block},
\end{equation}
with
\begin{subequations}
  \begin{eqnarray}
    \mathbf{A} &\equiv&     \begin{pmatrix}
      \breve{M}^b_u - Q_{22}^b      & G^b     \\
      Q_{32}^b+M_3^\rho M_3^{-1}D^b & M_3^\Pi
    \end{pmatrix},\label{eqn:hybrid-A}\\
    \mathbf{K} &\equiv& \left[K,0\right]^T,\label{eqn:hybrid-K}\\
    \mathbf{C} &\equiv& \left[C,0\right]^T`,\label{eqn:hybrid-C}\\
    \widetilde{\mathbf{X}}' &\equiv& \left[\widetilde{u}'^b,\widetilde{\Pi'}\right]^T,\label{eqn:hybrid-X}\\
    \vec{\mathcal{R}}_X &\equiv& \left[\mathcal{R}_u^b,\overline{\mathcal{R}}_\Pi\right]^T,\label{eqn:hybrid-Rx}
  \end{eqnarray}
\end{subequations}
where bold operators and vectors indicate objects that act on a combination of function spaces. Eliminating $\widetilde{\mathbf{X}}'$ from \eqref{eqn:gh-fem-hybrid-block} results in an equation for the Lagrange multipliers $\widetilde{\lambda}$
\begin{equation}
  S\widetilde{\lambda} = \mathcal{B}_\lambda\label{eqn:hybrid_equation}
\end{equation}
with
\begin{eqnarray}
  S &\equiv& \mathbf{C}^T\mathbf{A}^{-1}\mathbf{K}\qquad\text{and}\label{eqn:hybrid-S}\\
  \mathcal{B}_\lambda &\equiv& \mathbf{C}^T\mathbf{A}^{-1}\vec{\mathcal{R}}_X.\label{eqn:hybrid-rhs}
\end{eqnarray}
Since $\mathbf{A}^{-1}$ is block-diagonal and only couples unknown within a single grid cell, the Schur complement matrix $S$ in Eq. \eqref{eqn:hybrid-S} is (block-) sparse and can be applied cheaply while iteratively solving Eq. \eqref{eqn:hybrid_equation} with a Krylov subspace method. Observe further that instead of iterating in the pressure-velocity space as in Eq. \eqref{eqn:2x2system}, we now only need to iterate in the smaller trace space: at lowest order, we need to store on average three trace unknowns per cell (one shared trace unknown for each of the six facets). The right-hand side $\mathcal{B}_\lambda$ in Eq. \eqref{eqn:hybrid-rhs}, which can be interpreted as the matrix-representation of a one-form in the dual space $\mathbb{W}_2^{\text{tr}*}$, can be constructed using purely local operations. Once the solution $\widetilde{\lambda}$ of Eq. \eqref{eqn:hybrid_equation} is known, the Exner pressure $\widetilde{\Pi}'$ and (broken) velocity increment $\widetilde{u}'^b$ can be obtained via the local operation
\begin{equation}
  \widetilde{\mathbf{X}}' = \mathbf{A}^{-1}\left(\vec{\mathcal{R}}_X - \mathbf{K}\widetilde{\lambda}\right).\label{eqn:hybrid-recovery}
\end{equation}
\subsection{Non-nested multigrid}\label{sec:nonnested_multigrid}
To solve Eq. \eqref{eqn:hybrid_equation} we use a non-nested multigrid algorithm similar to that described in \cite{Gopalakrishnan2009}. The central idea is the following: as for any multigrid algorithm, first a small number of smoother iterations are applied to reduce high-frequency error components. Since the Lagrange multiplier field $\lambda\in\mathbb{W}_2^{\text{tr}}$ approximates the Exner pressure increment, we then restrict the residual to the pressure space $\mathbb{W}_3$ and solve the resulting coarse level equation with the pressure multigrid algorithm described in Section \ref{sec:approximate_schur_multigrid}, before prolongating the coarse level correction back to $\mathbb{W}_2^{\text{tr}}$. The same algorithm has been proposed for the mixed Poisson problem in \cite{Gopalakrishnan2009}, with the crucial difference that the authors of \cite{Gopalakrishnan2009} employ a conforming coarse level space instead of the discontinuous space used in this work. We begin by discussing an efficient smoother algorithm for Eq. \eqref{eqn:hybrid_equation}.
\paragraph{Trace space smoother}
As for the pressure multigrid algorithm in Section \ref{sec:approximate_schur_multigrid}, the smoother in the trace space $\mathbb{W}_2^{\text{tr}}$ needs to take account of the strong anisotropy caused by the high aspect ratio of the grid ($\Delta z\ll\Delta x$).  First note that the trace space $\mathbb{W}_2^{\text{tr}}$ can be written as the direct sum $\mathbb{W}_2^{\text{tr}} = \mathbb{W}_2^{\text{tr,h}}\oplus \mathbb{W}_2^{\text{tr,z}}$. As shown in Figure \ref{fig:function_spaces}, the degrees of freedom of $\mathbb{W}_2^{\text{tr,h}}$ are associated with vertical facets (aligned parallel to the vertical direction $\widehat{\vec{z}}$) while the degrees of freedom of $\mathbb{W}_2^{\text{tr,z}}$ are associated with horizontal facets (aligned perpendicular to the vertical direction $\widehat{\vec{z}}$). Similar to $\mathbb{W}_2^{\text{z}}$, the function space $\mathbb{W}_2^{\text{tr,z}}$ is continuous in the vertical but discontinuous in the horizontal direction, while $\mathbb{W}_2^{\text{tr,h}}$ is horizontally continuous and vertically discontinuous.

The decomposition of $\mathbb{W}_2^{\text{tr}}$ implies that the dof-vector $\widetilde{\lambda}=(\widetilde{\lambda}^{\text{h}},\widetilde{\lambda}^{\text{z}})$ can be partitioned such that the unknowns associated with $\mathbb{W}_2^{\text{tr,h}}$ are collected in the vector $\widetilde{\lambda}^{\text{h}}$, while the unknowns associated with $\mathbb{W}_2^{\text{tr,z}}$ are stored in $\widetilde{\lambda}^{\text{z}}$. Further, since $\mathbb{W}_2^{\text{tr,z}}$ is discontinuous in the horizontal direction, we can write $\widetilde{\lambda}^{\text{z}}=(\widetilde{\lambda}^{\text{z}}_1,\widetilde{\lambda}^{\text{z}}_2,\dots,\widetilde{\lambda}^{\text{z}}_{n_\text{col}})$ where $\widetilde{\lambda}^{\text{z}}_j$ is the dof-vector of all unknowns associated with a particular column $j$. With this ordering, the matrix $S$ in Eq. \eqref{eqn:hybrid_equation} has the following block-structure:
\begin{equation}
  S = \begin{pmatrix}
    S^{\text{hh}} & S^{\text{hz}} \\
    S^{\text{zh}} & S^{\text{zz}}
  \end{pmatrix}
  \qquad\text{with}\qquad
  S^{\text{zz}} = \text{blockdiag}\left(S^{\text{zz}}_1,S^{\text{zz}}_2,\dots,S^{\text{zz}}_{n_\text{col}}\right)
\end{equation}
Due to the strong vertical anisotropy, the entries of the matrix $S^{\text{zz}}$ are much larger than all other entries in $S$. Furthermore, the block-diagonal matrix $S^{\text{zz}}$ can be inverted by independently solving a tridiagonal system in each vertical column of the grid.

The smoother is written down explicitly in Algorithm \ref{alg:hybrid_linesmoother}, and has a very similar structure to the block-Jacobi method in Eq. \eqref{eqn:pressure_block_jacobi} that is used as a smoother for the pressure multigrid algorithm in Section \ref{sec:approximate_schur_multigrid}. While we use a block-Jacobi iteration for the degrees of freedom associated with $\mathbb{W}_2^{\text{tr,z}}$, a simple point Jacobi procedure is used for the unknowns associated with $\mathbb{W}_2^{\text{tr,h}}$. Recall that solving a system of the form
\begin{equation}
  S_j^{\text{zz}}\widetilde{\delta\lambda}_j=(\mathcal{R}_\lambda^{\text{z}})_j
  \label{eqn:banded_system}
\end{equation}
in line 5 of Algorithm \ref{alg:hybrid_linesmoother} with the Thomas algorithm
\cite{Press2007} will incur a cost of $\mathcal{O}(L)$ where $L$ is the number of vertical layers of the grid (at higher order $S^{\text{zz}}_j$ is a banded matrix with bandwidth $b$, which can be solved in $\mathcal{O}(bL)$ time if we ignore the initial overhead of the banded factorisation).

\begin{algorithm}
  \caption{$\textsf{LineSmooth}(\widetilde{\lambda};S,\mathcal{B}_\lambda;\omega,n_{\text{smooth}})$: update $\widetilde{\lambda}$ with $n_{\text{smooth}}$ (block-) Jacobi iterations with relaxation factor $\omega$}\label{alg:hybrid_linesmoother}
  \begin{algorithmic}[1]
    \For{$k=1,2,\dots,n_{\text{smooth}}$}
    \State Compute $\mathcal{R}_\lambda\mapsfrom\mathcal{B}_\lambda-S\widetilde{\lambda}$
    \State Partition $\mathcal{R}_\lambda=(\mathcal{R}_\lambda^{\text{h}},\mathcal{R}_\lambda^{\text{z}})$ and $\mathcal{R}_\lambda^{\text{z}}=((\mathcal{R}_\lambda^{\text{z}})_1,(\mathcal{R}_\lambda^{\text{z}})_2,\dots,(\mathcal{R}_\lambda^{\text{z}})_{n_\text{col}})$ in the same way as $\widetilde{\lambda}$
    \For{$j=1,2,\dots,n_{\text{col}}$}
    \State Solve the tridiagonal system $S_j^{\text{zz}}\widetilde{\delta\lambda}_j=(\mathcal{R}_\lambda^{\text{z}})_j$ for $\widetilde{\delta\lambda}_j$
    \State Update $\widetilde{\lambda}^{\text{z}}_j\mapsfrom \widetilde{\lambda}^{\text{z}}_j-\omega \widetilde{\delta\lambda}_j$
    \EndFor
    \State Update $\widetilde{\lambda}^{\text{h}}\mapsfrom \widetilde{\lambda}^{\text{h}}-\omega \left(\text{diag}(S^{\text{hh}})\right)^{-1} \mathcal{R}_\lambda^{\text{h}}$
    \EndFor
  \end{algorithmic}
\end{algorithm}

Of course, the smoother in Algorithm \ref{alg:hybrid_linesmoother} could be used directly to precondition a Krylov-subspace method for the solution of $S\widetilde{\lambda}=\mathcal{B}_\lambda$, and in our numerical results in Section \ref{sec:results} we also investigate this setup as a simpler reference method. As expected for a single-level preconditioner, the number of Krylov iterations turns out to be very large, which leads to a larger overall solution time. The hierarchical algorithm that we describe next overcomes this issue.
\paragraph{Non-nested two level algorithm}
A single iteration of the non-nested two-level method is described in Algorithm \ref{alg:multigrid}, where $\textsc{LineSmooth}(\widetilde{\lambda};S,\mathcal{B}_{\lambda},\omega,n_{\text{smooth}})$ corresponds to $n_{\text{smooth}}$ applications of the linesmoother in Algorithm \ref{alg:hybrid_linesmoother}. Crucially, the operator $H$ in the coarse level system
\begin{equation}
  H \widetilde{\delta\Pi} = \overline{\mathcal{B}}
  \label{eqn:coarse_level_system}
\end{equation}
in line 6 of Algorithm \ref{alg:multigrid} is the operator defined in Eq. \eqref{eqn:elliptic_pressure_system}, and we can therefore solve Eq. \eqref{eqn:coarse_level_system} with the pressure multigrid algorithm from Section \ref{sec:approximate_schur_multigrid}.
\begin{algorithm}
  \caption{Non-nested two-level cycle for approximately solving $S\widetilde{\lambda}=\mathcal{B}_\lambda$}\label{alg:multigrid}
  \begin{algorithmic}[1]
    \State Set $\widetilde{\lambda} \mapsto 0$
    \State Presmooth: $\textsc{LineSmooth}(\widetilde{\lambda};S,\mathcal{B}_\lambda;\omega,n_{\text{presmooth}})$ (see Algorithm \ref{alg:hybrid_linesmoother})
    \State Compute residual: $\mathcal{R}_\lambda\mapsfrom\mathcal{B}_\lambda-S\widetilde{\lambda}$
    \State Restrict residual: $\mathcal{B}\mapsfrom\textsc{Restrict}\left(\mathcal{R}_\lambda\right)$
    \State Scale right-hand side: $\overline{\mathcal{B}}\mapsfrom\Gamma \mathcal{B}$ (see Eq. \eqref{eqn:scaling_matrix})
    \State Solve pressure system $H \widetilde{\delta\Pi} = \overline{\mathcal{B}}$ for $\widetilde{\delta\Pi}$.
    \State Prolongate correction and update solution: $\widetilde{\lambda}\mapsfrom\widetilde{\lambda}+\textsc{Prolong}\left(\widetilde{\delta\Pi}\right)$
    \State Postsmooth: $\textsc{LineSmooth}(\widetilde{\lambda};S,\mathcal{B}_\lambda;\omega,n_{\text{postsmooth}})$ (see Algorithm \ref{alg:hybrid_linesmoother})
  \end{algorithmic}
\end{algorithm}
\paragraph{Intergrid operators}
To prolongate the solution from the pressure space $\mathbb{W}_3$ to the trace space $\mathbb{W}_2^{\text{tr}}$ we use the fact that $\lambda\in\mathbb{W}_2^{\text{tr}}$ is an approximation of the Exner pressure increment $\Pi'\in \mathbb{W}_3$. Hence, the natural prolongation $\mathfrak{P}:\mathbb{W}_3\rightarrow \mathbb{W}_2^{\text{tr}}$ is the averaging procedure
\begin{equation}
  \mathfrak{P}:\Pi'\mapsto \lambda=\favg{\Pi'}.
\end{equation}
The corresponding linear operator acting on the dof-vector $\widetilde{\Pi}'$ can be written as $\textsf{Prolong}\left(\widetilde{\Pi}'\right) \equiv P \widetilde{\Pi}'$. At lowest order, the application of the matrix $P$ simply corresponds to averaging the values of the unknowns on neighbouring cells to obtain the trace-unknown of the facet between those two cells. Restricting the dof-vector $\widetilde{\lambda}$ of a field in the trace-space $\mathbb{W}_2^{\text{tr}}$ to the dof-vector of a field in the pressure space $\mathbb{W}_3$ then corresponds to the multiplication with the transpose matrix, i.e. $\textsf{Restrict}\left(\widetilde{\lambda}\right) \equiv P^T\widetilde{\lambda}$.
\paragraph{Coarse level operator}
Given a fine level matrix $S$, a prolongation matrix $P$ and a restriction matrix $R=P^T$, the canonical way of constructing a suitable coarse level operator is to form the Galerkin product $H_\text{G}=P^T S P$. This leads to the coarse level equation $H_G\widetilde{\delta \Pi}=\mathcal{B}=R^T\mathcal{R}_\lambda$. Unfortunately, the triple matrix-product required in the construction of $H_G$ is expensive and $H_G$ has a large stencil. To address these issues we use a different approach here. We know that $H$ as defined in Eq. \eqref{eqn:elliptic_pressure_system} (which only contains couplings between adjacent cells) is a good approximation to $H_G$, in the sense that the action of $H_G$ on smooth fields is similar to the action of $H$, up to some scaling factor $\Gamma$, which appears in line 5 of Algorithm \ref{alg:multigrid}. More precisely,
\begin{equation}
  \Gamma H_G \widetilde{\Pi}_{\text{smooth}} \approx H \widetilde{\Pi}_{\text{smooth}}
\end{equation}
where $\Gamma$ is a diagonal matrix and $\widetilde{\Pi}_{\text{smooth}}$ is the dof-vector of a $\mathbb{W}_3$ field that varies slowly in space. To find the entries of $\Gamma$, we consider the constant field $\widetilde{\Pi}_{\text{const}}$ with $(\widetilde{\Pi}_{\text{const}})_j=1$ for all $j$. Then $\Gamma_{jj}$ is given by the ratio of the $j$-th row-sums of $H$ and $H_G$:
\begin{equation}
  \Gamma_{jj} =
  \frac{\sum_{k}\left(H\right)_{jk}}{\sum_{k}\left(H_G\right)_{jk}} =
  \frac{(H \widetilde{\Pi}_{\text{const}})_j}{(H_G \widetilde{\Pi}_{\text{const}})_j}
  =
  \frac{(H \widetilde{\Pi}_{\text{const}})_j}{(P^T SP \widetilde{\Pi}_{\text{const}})_j}.
  \label{eqn:scaling_matrix}
\end{equation}
Evaluating the expressions $H \widetilde{\Pi}_{\text{const}}$ and $P^TSP \widetilde{\Pi}_{\text{const}}$ that are needed in the numerator and denominator of the rightmost fraction in Eq. \eqref{eqn:scaling_matrix} requires only multiplications by the sparse matrices $S$, $H$, $P$ and $P^T$.

The coarse level problem $H\widetilde{\delta H}=\overline{\mathcal{B}}$ in line 6 of Algorithm \ref{alg:multigrid} is solved with the pressure multigrid algorithm described in Section \ref{sec:approximate_schur_multigrid}.
\section{Results}\label{sec:results}
\subsection{Implementation}\label{sec:implementation}
The non-nested multigrid solver described in Section \ref{sec:nonnested_multigrid} has been implemented in the full LFRic model, which uses a separation of concerns philosophy to orchestrate the efficient parallel execution of computational kernels \cite{adams2019lfric}. In our case, these kernels correspond, for example, to the application of the linear operator $S$ defined in Eq. \eqref{eqn:hybrid-S} in a vertical column, or a single trididiagonal solve in Algorithm \ref{alg:hybrid_linesmoother}. From the  discussion in Section \ref{sec:methods} it should be evident that the solver architecture is highly non-trivial, and it is crucial to have a simple yet robust mechanism for independently testing and swapping different solver components. This allows a comparison to be made to alternative approaches, such as the approximate Schur complement method in \cite{Maynard2020}, to ultimately identify the most promising solver configuration. To realise this, a hierarchy of Fortran 2003 derived types has been developed that conform to common interfaces: following the same modular design as other libraries \cite{petsc-user-ref,petsc-web-page,petsc-efficient,blatt2006iterative}, the ``Solver API'' in LFRic \cite{Maynard2020,adams2019lfric} allows the implementation of iterative solvers, linear operators and preconditioners which can be plugged together by the algorithm developer. Further details on the modular implementation of the solver algorithms discussed in this paper can be found in Appendix \ref{sec:appendix_code_architecture}.
\subsection{Solver configurations}\label{sec:solver_configurations}
This paper introduces two innovations: the solution of the linear system in Eqs. \eqref{eqn:gh-u} - \eqref{eqn:gh-pi} with the help of a hybridised discretisation and  efficient preconditioning of the hybridised system in Eq. \eqref{eqn:hybrid_equation} with a non-nested multigrid algorithm. To assess the impact of each of these changes on model performance, we compare three different solver setups:
\begin{description}
  \item [Configuration 1 (Pressure multigrid):] The current non-hybridised solver introduced in \cite{Maynard2020}, preconditioning the mixed system in Eq. \eqref{eqn:2x2system} with the approximate Schur complement algorithm that solves the pressure system with multigrid as described in Section \ref{sec:approximate_schur_multigrid}. This configuration can be considered as a baseline for our comparisons.
  \item [Configuration 2 (Hybridised single-level):] The hybridised solver as described in Section \ref{sec:hybridisation}, preconditioning the trace system in Eq. \eqref{eqn:hybrid_equation} with a few iterations of the block-Jacobi smoother in Algorithm \ref{alg:hybrid_linesmoother}.
  \item [Configuration 3 (Hybridised multigrid):] The hybridised solver as described in Section \ref{sec:hybridisation}, preconditioning the trace system in Eq. \eqref{eqn:hybrid_equation} with the non-nested multigrid algorithm in Algorithm \ref{alg:multigrid}.
\end{description}
The gains in solver performance achieved by hybridising the equations can be assesed by comparing configurations 1 and 3, using robust multigrid solvers for both discretisations. However, given the complexity of the non-nested multigrid algorithm in Algorithm \ref{alg:multigrid}, it is worth assessing whether the hybridised solver, using a much simpler single-level preconditioner, is competitive compared to the non-hybridised solver. As in \cite{Maynard2020}, we therefore compare our hybridised multigrid preconditioner (Configuration 3) to the block-Jacobi preconditioner (Configuration 2).

For the pressure multigrid preconditioner (Configuration 1), the mixed system in Eq. \eqref{eqn:2x2system} is solved with BiCGStab \cite{VanderVorst1992,Saad2003}, reducing the two-norm of the residual by six orders of magnitude. The same BiCGStab iteration and relative residual reduction is used to solve the system in Eq. \eqref{eqn:hybrid_equation} for the two hybridised solvers (Configurations 2 and 3).

For the hybridised single-level preconditioner (Configuration 2), we can apply different numbers of smoothing steps, with a trade-off between the total number of BiCGStab iterations and the time per iteration. Thus, we consider three different setups, with $n_{\text{smooth}} = $ 1, 2 and 3 in Algorithm \ref{alg:hybrid_linesmoother}. Note that using $n_{\text{smooth}} = 3$ then uses the equivalent number of smoothing steps as the hybridised multigrid setup (Configuration 3), which uses $n_{\text{presmooth}} = 1$ and $n_{\text{postsmooth}} = 2$.

In the non-nested two-level method in Algorithm \ref{alg:multigrid}, the coarse level pressure system in Eq. \eqref{eqn:coarse_level_system} can be solved in different ways. Since Algorithm \ref{alg:multigrid} is used as a preconditioner for a BiCGStab iteration and the coarse level solution $\widetilde{\delta H}$ is an approximation to the solution of the residual equation, it is usually sufficient to only solve it to some loose tolerance, which will lead to better overall performance. To explore this, we consider three different setups for the coarse level solve:
\begin{description}
  \item[Configuration 3a (exact coarse solve):] Solve Eq. \eqref{eqn:coarse_level_system} to a very tight tolerance, reducing the relative residual norm below $10^{-14}$.
  \item[Configuration 3b (approximate coarse solve):] Solve Eq. \eqref{eqn:coarse_level_system} to a loose tolerance, reducing the residual norm by two orders of magnitude only.
  \item[Configuration 3c (pressure multigrid V-cycle coarse solve):] Apply one V-cycle of the pressure multigrid algorithm from Section \ref{sec:approximate_schur_multigrid} to approximately solve Eq. \eqref{eqn:coarse_level_system} as cheaply as possible.
\end{description}
In Configurations 3a and 3b, the coarse level solver is BiCGStab, preconditioned with the existing pressure multigrid algorithm from Section \ref{sec:approximate_schur_multigrid}. Configuration 3a, which essentially solves the coarse level problem in Eq. \eqref{eqn:coarse_level_system} to machine precision, is useful to debug any issues with Algorithm \ref{alg:multigrid}. It is worth recalling that we used similar setups in \cite{Maynard2020} to verify that solving the pressure correction there with a single multigrid V-cyle gives the best overall performance.

For reference, Table \ref{tab:solver_configurations} summarises the different solver configurations that were used for the numerical experiments below.
\def\arraystretch{1.25}
\begin{table}[H]
  \begin{center}
    \rowcolors{2}{white}{black!10}
    \begin{tabular}{|c|c|c|c|}
      \hline
      Config                                    & mixed solve+precon                               & inner solve+precon                      & coarse solve+precon              \\
      \hline
      1                                         & $\textsc{BiCGStab}(10^{-6})+\text{approx Schur}$ &
      $\textsc{pre-only}+\textsc{MG}$           & ---                                                                                                                           \\
      2                                         & pre-only+hybridisation                           &
      $\textsc{BiCGStab}(10^{-6})$ + LineSmooth & ---                                                                                                                           \\
      3a                                        & pre-only+hybridisation                           & $\textsc{BiCGStab}(10^{-6})$+non-nested & $\textsc{BiCGStab}(10^{-14})$+MG \\
      3b                                        & pre-only+hybridisation                           & $\textsc{BiCGStab}(10^{-6})$+non-nested & $\textsc{BiCGStab}(10^{-2})$+MG  \\
      3c                                        & pre-only+hybridisation                           &
      $\textsc{BiCGStab}(10^{-6})$+non-nested   & $\textsc{pre-only}$+MG                                                                                                        \\\hline
    \end{tabular}
  \end{center}
  \caption{Summary of solver configurations. Using the ``pre-only'' solve corresponds to a single application of the preconditioner. ``LineSmooth'' is the block-Jacobi smoother in Algorithm \ref{alg:hybrid_linesmoother}, ``non-nested'' refers to Algorithm \ref{alg:multigrid} whereas ``MG'' stands for the pressure multigrid algorithm described in Section \ref{sec:approximate_schur_multigrid}.}
  \label{tab:solver_configurations}
\end{table}
\subsection{Model setup}
To obtain the results which follow, each configuration in Table \ref{tab:solver_configurations} is run for 12 timesteps with a horizontal wave Courant number of approximately 12; this mimics what would be used in operational settings. The pressure multigrid preconditioner uses 4 multigrid levels and 1 pre-smoothing and 2 post-smoothing steps with a relaxation factor of $\omega_{\text{pressure}}=0.9$; this setup is used both in Configuration 1 and when pressure multigrid is used as a coarse level solver in Configuration 3. The same number of pre- and post-smoothing steps are also used for the non-nested two-level algorithm in Algorithm \ref{alg:multigrid}. However, in this algorithm and for the smoother that preconditions the single-level method in Configuration 2, we set $\omega=0.6$. The dependence of solver performance on the different parameters is explored in Appendix \ref{sec:solver_parameters}, where it is demonstrated that the chosen values give good performance in general.  As discussed in \cite{Maynard2020} and initially observed in \cite{Sandbach2015}, it is sufficient to use a small number of multigrid levels. This is because for the chosen Courant numbers the condition number of Eq. \eqref{eqn:coarse_level_system} is reduced to $\approx 1$ on the coarsest  multigrid level, independent of the horizontal resolution. We perform all experiments in parallel, using the Met Office Cray XC40 supercomputer which uses the Aries interconnect. Each node comprises dual-socket, 18-core Broadwell Intel Xeon processors, that is, 36 CPU cores per node. The model was compiled with the Intel 17 Fortran compiler.

We simulate the full compressible Euler equations for a perfect gas, as set out in Section \ref{sec:model_eqns}. The equations are solved in a spherical domain without orography, which is covered with a cubed sphere mesh that contains $6\times 96 \times 96$ vertical columns (this mesh is referred to as ``C96'' below). Each column is subdivided into 30 vertical layers with a grid spacing which increases quadratically with height above ground. Overall, the mesh contains $1,658,880$ grid cells and the average horizontal grid spacing is $\approx 96~\text{km}$. The model timestep size is set to $1~\text{hour}$, resulting in a horizontal acoustic Courant number of $\approx 12$. The radius of the earth is assumed to be $6371.229~ \text{km}$; the vertical lid is fixed at $30~\text{km}$ above ground. All calculations were carried out in double precision (64bit floating point) arithmetic.
\subsection{Single-node Performance}\label{sec:results_single_node}
Two key metrics are considered to assess the performance of each solver configuration:
\begin{enumerate}
  \item The average time per model timestep.
  \item The convergence history of the preconditioned iterative solvers (for Eq. \eqref{eqn:2x2system} in Configuration 1 and Eq. \eqref{eqn:hybrid_equation} in Configurations 2 and 3).
\end{enumerate}
While the ultimate goal is to reduce the total runtime, the convergence history provides valuable insights into the algorithmic performance of the solvers and can highlight the potential for further optimisation. For multigrid we typically expect the residual to be reduced by one order of magnitude per iteration, while (as already demonstrated in \cite{Maynard2020}) convergence is much slower for single-level preconditioners; this is confirmed by comparing Configurations 2 and 3 in the following results.

\begin{figure}[!ht]
  \centering
  \includegraphics[width=0.8\textwidth]{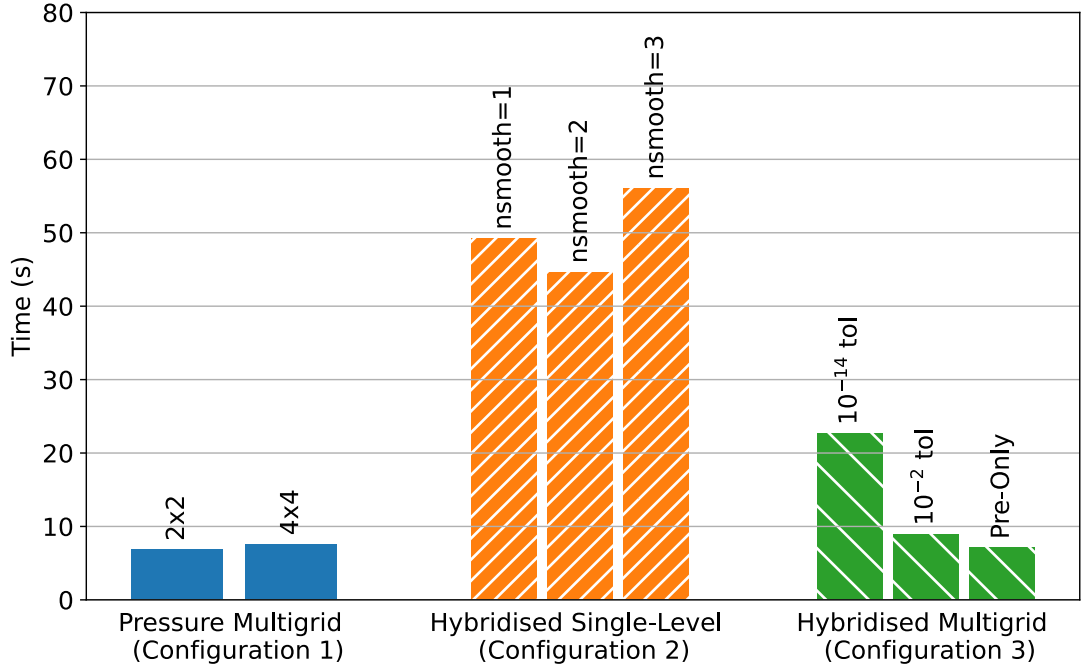}
  \caption{Average time per timestep for the full compressible Euler equations in LFRic.}
  \label{fig:time_per_ts}
\end{figure}

Figure \ref{fig:time_per_ts} shows the average time per timestep for all the considered solver configurations, running the model on a single node (36 cores) of the Met Office Cray XC40 supercomputer and averaging over all 12 model timesteps. The pressure multigrid setup (Configuration 1) has been further separated into a method which iterates over the $4\times4$ system as in \cite{Maynard2020} and the new method described here, which eliminates potential temperature and density as described in Sections \ref{sec:model_eqns} and \ref{sec:mixed_finite_element_discretisation} to obtain a $2\times2$ system in Eq. \eqref{eqn:2x2system}. While the old method takes $7.60\text{s}$ per timestep, iterating over the $2\times 2$ system reduces this to $6.83\text{s}$, which demonstrates that eliminating potential density and temperature before the iteration improves performance by around $10\%$. For the rest of this paper only the $2\times2$ system will be considered for Configuration 1.
\noindent
We observe that the non-hybridised pressure multigrid solver (Configuration 1) gives the best performance (6.82~\text{s}), but is closely followed by our hybridised multigrid solver (Configuration 3c), which is around $4\%$ slower (7.08~\text{s}). Solving the coarse level system in the non-nested two-level solver to a tighter tolerance increases the runtime to 22.63~\text{s} for Configuration 3a and 8.92~\text{s} for Configuration 3b. One striking feature of Figure \ref{fig:time_per_ts} is that the hybridised single-level solver (Configuration 2) gives the worst runtime overall -- 49.29~\text{s}, 44.64~\text{s}, and 56.06~\text{s} for $n_{\text{smooth}} = 1$, 2, and 3 respectively. In other words, replacing the linesmoother preconditioner in the hybridised solver with the non-nested multigrid algorithm reduces the runtime by a factor of more than $6\times$ and only with this change does our hybridised solver become competitive with the pressure multigrid setup in \cite{Maynard2020}.

\begin{figure}[!ht]
  \centering
  \includegraphics[width=\textwidth]{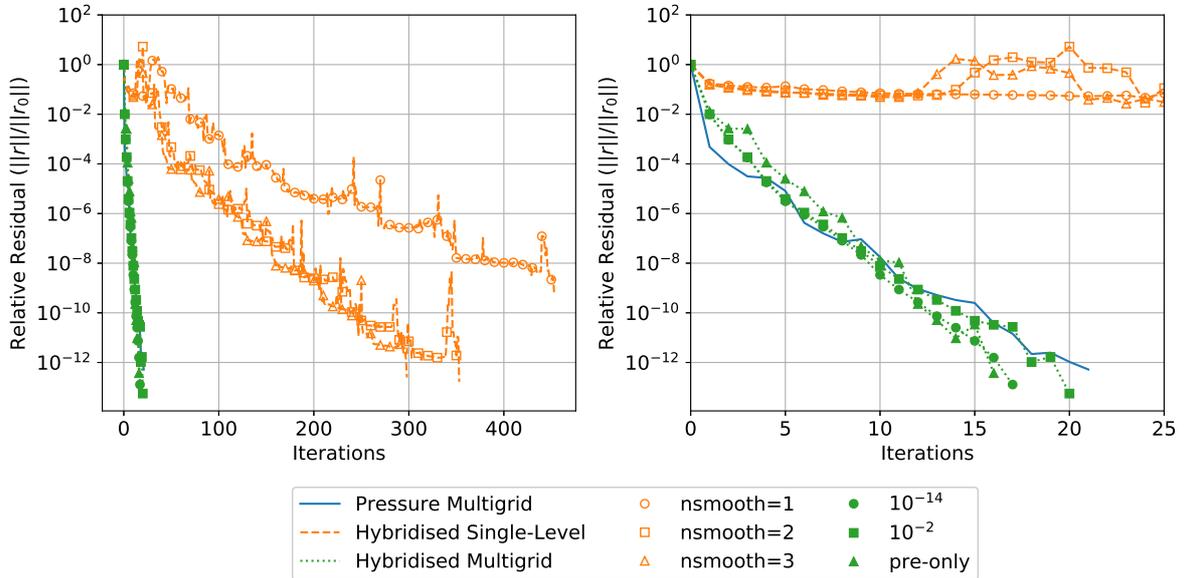}
  \caption{Convergence history of the iterative solvers (left), with a zoom to the first 20 iterations (right).}
  \label{fig:parallel6_1e-12_convergence_history}
\end{figure}

Figure \ref{fig:parallel6_1e-12_convergence_history} shows the convergence histories for the relative residual $||r||/||r_0||$.
\noindent
We observe that the convergence rates for the three variations of the non-nested two-level preconditioner (Configuration 3) are comparable to that observed for the pressure multigrid solver (Configuration 1). In contrast, the convergence rate of the hybridised single-level solver (Configuration 2) is prohibitively slow and this is reflected in the time to solution in Figure \ref{fig:time_per_ts}: more than 400 iterations are required to reduce the residual by eight orders of magnitude for $n_{\text{smooth}} = 1$, and in fact  Configuration 2 with $n_{\text{smooth}} = 1$ fails to reduce the residual by a relative factor $~10^{-9}$. In contrast, to reduce the residual by twelve orders of magnitude our hybridised multigrid solver (Configuration 3c) requires 16 iterations -- which is textbook multigrid performance -- and the pressure multigrid solver (Configuration 1) takes slightly longer to converge with 21 iterations.

The convergence history also explains the poor performance of Configurations 3a and 3b in Figure \ref{fig:time_per_ts}: Figure \ref{fig:parallel6_1e-12_convergence_history} demonstrates that the multigrid convergence rate is virtually independent of the coarse level solution method. This is readily explained by the fact that the coarse level equation in Eq. \eqref{eqn:coarse_level_system} is an approximation of Eq. \eqref{eqn:hybrid_equation}, so there is not much point in solving it to high accuracy. This mirrors the findings in \cite{Maynard2020}, where we showed that the best performance is achieved by approximately solving the pressure correction equation with a single multigrid V-cycle.

The convergence history for the hybridised single-level solver (Configuration 2) shows large spikes, with the residual increasing by more than one order of magnitude from one iteration to the next in some cases. In practice, this can cause serious issues when determining whether the solver has converged. It also potentially makes the runtime unpredictable, which is undesirable for operational forecast models that must run in a fixed time window. More importantly, the slow convergence of the single-level solver will likely cause serious issues when running the model in single precision floating point arithmetic -- the Krylov subspace vectors lose orthogonality in reduced precision calculations and the convergence will stall.

Comparing the different number of smoothing applications $n_{\text{smooth}}$ for the hybridised single-level solver (Configuration 2) illustrates that the linesmoother is effective in reducing the number of iterations to convergence. However, this of course results in a greater cost per BiCGStab iteration and thus the hybridised single-level solver is (as expected) not competitive with the pressure multigrid (Configuration 1) and hybridised multigrid (Configuration 3) solvers. We also observe that there is a limit to what the smoother can do: the convergence improves to a much lesser extent when increasing from $n_{\text{smooth}} = 2$ to $n_{\text{smooth}} = 3$, which means that the runtime using $n_{\text{smooth}} = 3$ is greatly increased, and is in fact very similar to the runtime when using $n_{\text{smooth}} = 1$.

Interestingly, in contrast to what was observed for a simplified gravity wave model (not shown here) and the Krylov subspace solvers with single-level preconditioners in the ENDGame model (also not shown here), in Figure \ref{fig:parallel6_1e-12_convergence_history} the residual for the single-level method (Configuration 2) does not drop rapidly in the first few iterations. Hence, even if the residual is reduced by less than six orders of magnitude, the hybridised multigrid preconditioner (Configuration 3c) will likely retain its advantage over the single-level method (Configuration 2).

\subsection{Breakdown of runtimes}\label{sec:breakdown_times}
\begin{figure}[!ht]
  \centering
  \includegraphics[width=0.6\textwidth]{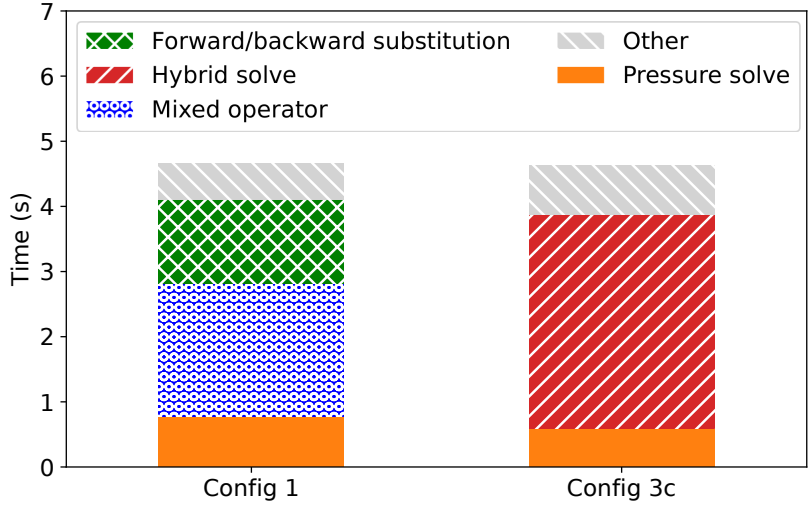}
  \caption{Bar chart illustrating the breakdown of time spent in the mixed solve.}
  \label{fig:timing_breakdown}
\end{figure}

To gain further insight into the different configurations, we also perform timings on key components of the solver. Only the timings from the pressure multigrid setup (Configuration 1) and our hybridised multigrid solver with the ``precondition only'' setup (Configuration 3c) are shown here due to the much greater cost of the other solver configurations. The results for these two configurations are presented in the bar chart in Figure \ref{fig:timing_breakdown}, where the total height of each bar represents the total time spent in the mixed solver; this is the average time it takes to solve Eq. \eqref{eqn:2x2system} with a suitably preconditioned Krylov subspace method. The figure shows the total time spent in all four linear solves that are required per timestep. Note that this is different to the times presented in Figure \ref{fig:time_per_ts} which are the times for executing an entire timestep (including, for example, the construction of the operators and non-linear residual calculation). Each bar is then broken down into key components, for example the time spent in solving the pressure correction equation. For Configuration 3c, ``Pressure solve'' refers to the time spent in the solution of the coarse level pressure system, whereas ``Hybrid solve'' measures the combined time spent in the application of the linesmoother (Algorithm \ref{alg:hybrid_linesmoother}) and trace operator $S$ in Eq. \eqref{eqn:hybrid-S}. Together, these times give a good estimate for the total time it takes to solve Eq. \eqref{eqn:hybrid_equation}. For Configuration 1 ``Mixed operator'' is the time spent in the application of the $2\times 2$ operator in Eq. \eqref{eqn:2x2system}, whereas ``Forward/backward substitution'' measures the construction of the right hand side for the pressure equation in Eq. \eqref{eqn:pressure_rhs} and the recovery of the velocity field from the pressure solve in Eq. \eqref{eqn:velocity_recovery}.

For the pressure multigrid setup (Configuration 1), the solution of the pressure correction system accounts for about 16~\% of the solve time, which is dominated by other parts of the mixed solver, such as in the application of the mixed operator and forward/backward substitution. The hybridised multigrid solver (Configuration 3c) spends a significant fraction of the total time in application of the hybridised operator and the linesmoother (namely in performing the hybrid solve), while the pressure multigrid is also important here.

As a comparison, it is worth noting that the hybridised single-level solver with $n_{\text{smooth}} = 1$ (Configuration 2a) spends around 46.8~s per timestep in the mixed solver, 72~\% of which is in the application of the hybridised operator and the linesmoother, which are called frequently in the preconditioned Krylov solve. As above, it is clear that the slow convergence rate greatly increases the time spent in the mixed solver -- the pressure multigrid setup (Configuration 1) and the hybridised multigrid solver with precondition only (Configuration 3c) spend 10$\times$ less time in solving the linear system in Eq. \eqref{eqn:2x2system} (around 4.7~s and 4.6~s respectively).

For reference, we also present a breakdown of the total time (integrated over the entire run) and the time per call for key solver components in Table \ref{tab:time_in_components}.
\def\arraystretch{1.25}
\begin{table}[H]
  \begin{center}
    \rowcolors{4}{white}{black!10}
    \begin{tabular}{|l|S[table-format=1.2,parse-numbers=false]|S[table-format=1.5,parse-numbers=false]|S[table-format=3.2,parse-numbers=false]|S[table-format=1.4,parse-numbers=false]|S[table-format=2.2,parse-numbers=false]|S[table-format=1.5,parse-numbers=false]|}
      \hline
                                                           & \multicolumn{6}{c|}{Time [s]}                                                                                                                                                                \\\hline
                                                           & \multicolumn{2}{c|}{Config 1} & \multicolumn{2}{c|}{Config 2a} & \multicolumn{2}{c|}{Config 3c}                                                                                              \\\hline
      Component
                                                           & \multicolumn{1}{c|}{Total}    & \multicolumn{1}{c|}{per call}  & \multicolumn{1}{c|}{Total}     & \multicolumn{1}{c|}{per call} & \multicolumn{1}{c|}{Total} & \multicolumn{1}{c|}{per call} \\
      \hline
      pressure operator                                    & 5.13                          & 0.00048                        & {\textemdash}                  & {\textemdash}                 & 3.97                       & 0.00041                       \\
      approximate Schur complement preconditioner          & 8.90                          & 0.029                          & {\textemdash}                  & {\textemdash}                 & {\textemdash}              & {\textemdash}                 \\
      pressure multigrid V-cycle                           & 9.33                          & 0.010                          & {\textemdash}                  & {\textemdash}                 & 7.10                       & 0.087                         \\
      hybridised operator                                  & {\textemdash}                 & {\textemdash}                  & 284.40                         & 0.0085                        & 13.83                      & 0.0084                        \\
      hybridised forward/backward substitution             & {\textemdash}                 & {\textemdash}                  & 2.10                           & 0.044                         & 2.13                       & 0.044                         \\
      non-nested multigrid algorithm (inc. pressure solve) & {\textemdash}                 & {\textemdash}                  & {\textemdash}                  & {\textemdash}                 & 42.29                      & 0.052                         \\
      single-level preconditioner for hybridised system    & {\textemdash}                 & {\textemdash}                  & 131.33                         & 0.0039                        & 25.61                      & 0.016                         \\
      \hline
    \end{tabular}
  \end{center}
  \caption{Summary of the total time and time spent per call in several key components of the three solver configurations.}
  \label{tab:time_in_components}
\end{table}
\subsection{Parallel scaling}\label{sec:parallel_scaling}

The metrics considered above give a useful insight into how each solver performs per timestep, thus highlighting the algorithmic advantages of different methods. However, a key factor in choosing solvers is how they scale within massively parallel architectures.

To this end, we repeat the numerical experiments in Section \ref{sec:results_single_node} at scale to investigate the parallel performance of the two most promising solver configurations, namely the pressure multigrid setup (Configuration 1) and the hybridised multigrid solver (Configuration 3c). In all cases the timestep size is adjusted such that the horizontal acoustic Courant number is $\approx 12$ as for the experiments in Section \ref{sec:results_single_node}. For the weak scaling analysis the model is run on 6, 24, 96 and 384 nodes (recall that each node consists of 36 cores). The highest horizontal resolution that was considered is 12km on 13,824 cores. Details on the setup for these runs are collected in Table \ref{tab:setup_weak_scaling}.
\begin{table}
  \begin{center}
    \rowcolors{2}{white}{black!10}
    \begin{tabular}{|r|r|r|cp{1ex}r|r|r|}
      \hline
      \# nodes & \# cores & mesh & \multicolumn{3}{c|}{total \# grid cells} & resolution $\Delta x$ & timestep $\Delta t$                                  \\
      \hline
      6        & 216      & C96  & $6\times96\times96\times30$              & =                     & 1,658,880           & $96\text{km}$ & $3600\text{s}$ \\
      24       & 864      & C192 & $6\times192\times192\times30$            & =                     & 6,635,520           & $48\text{km}$ & $1800\text{s}$ \\
      96       & 3,456    & C384 & $6\times384\times384\times30$            & =                     & 26,542,080          & $24\text{km}$ & $900\text{s}$  \\
      384      & 13,824   & C768 & $6\times768\times768\times30$            & =                     & 106,168,320         & $12\text{km}$ & $450\text{s}$  \\
      \hline
    \end{tabular}
  \end{center}
  \caption{Setup of weak scaling runs.}
  \label{tab:setup_weak_scaling}
\end{table}
All runs use pure MPI parallelisation with 36 MPI regions per node, giving a local volume of $16^2$ columns (all with 30 vertical layers) per region for the weak scaling runs. It is anticipated that the future operational configuration of LFRic will run at a global resolution of around $5\text{km}$, which would correspond to a C1792 mesh with $6\times 1792\times 1792$ horizontal grid cells. Assuming that the number of columns per processor is kept fixed at $16^2$ (i.e. using the above weak scaling assumptions), this would require running the model on approximately $75,000$ cores. However, it will likely be necessary to increase the number of cores beyond this to (i) account for increases in the vertical resolution and to (ii) compensate for the larger number of timesteps that are necessary to keep the Courant number fixed. While one can only speculate on the size of the machine that will be required to run the full model in the future, the number of cores will likely be in the region of $10^5-10^6$.

\begin{figure}[!ht]
  \centering
  \subfloat[][]{\includegraphics[width=0.5\textwidth]{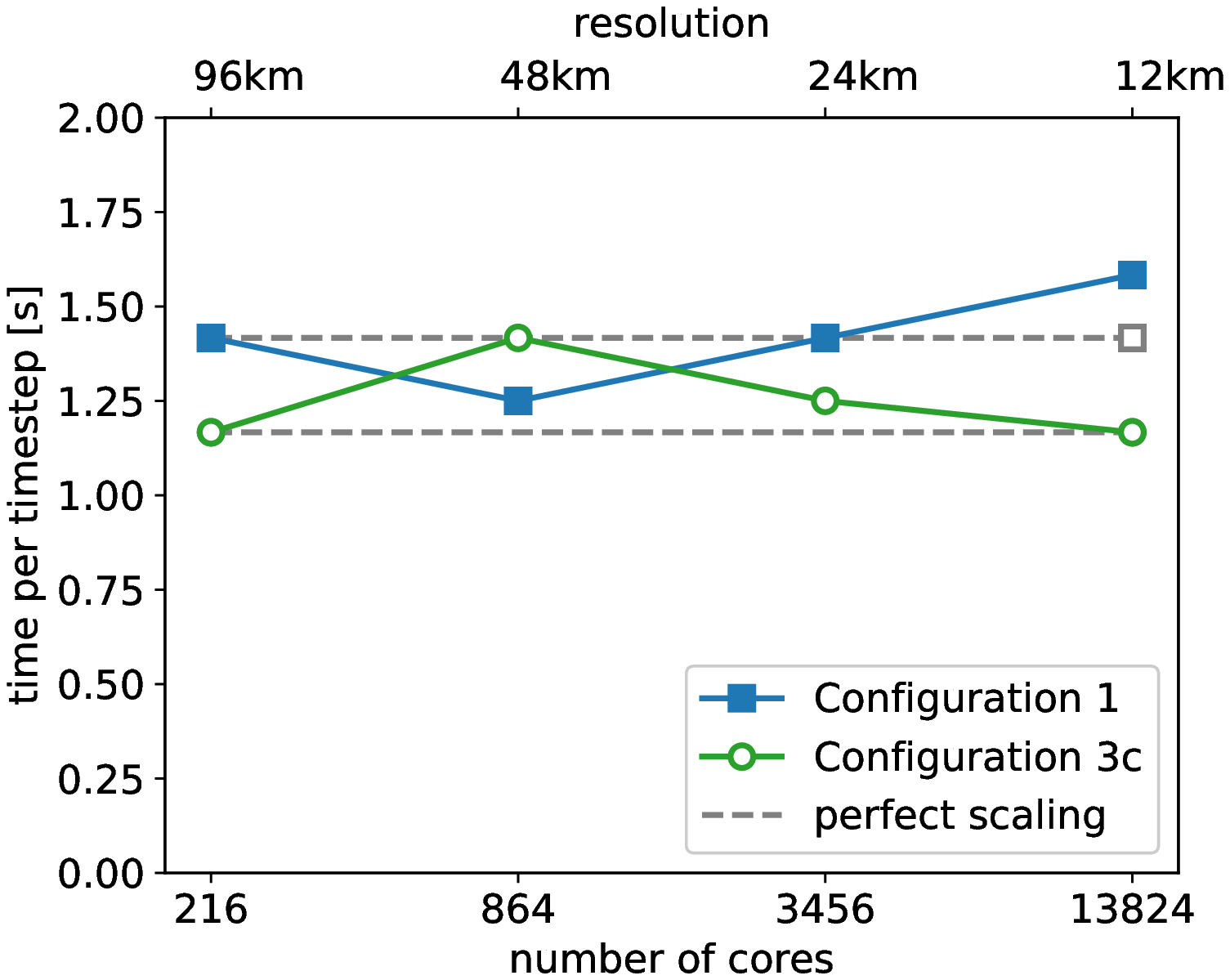}}
  \subfloat[][]{\includegraphics[width=0.5\textwidth]{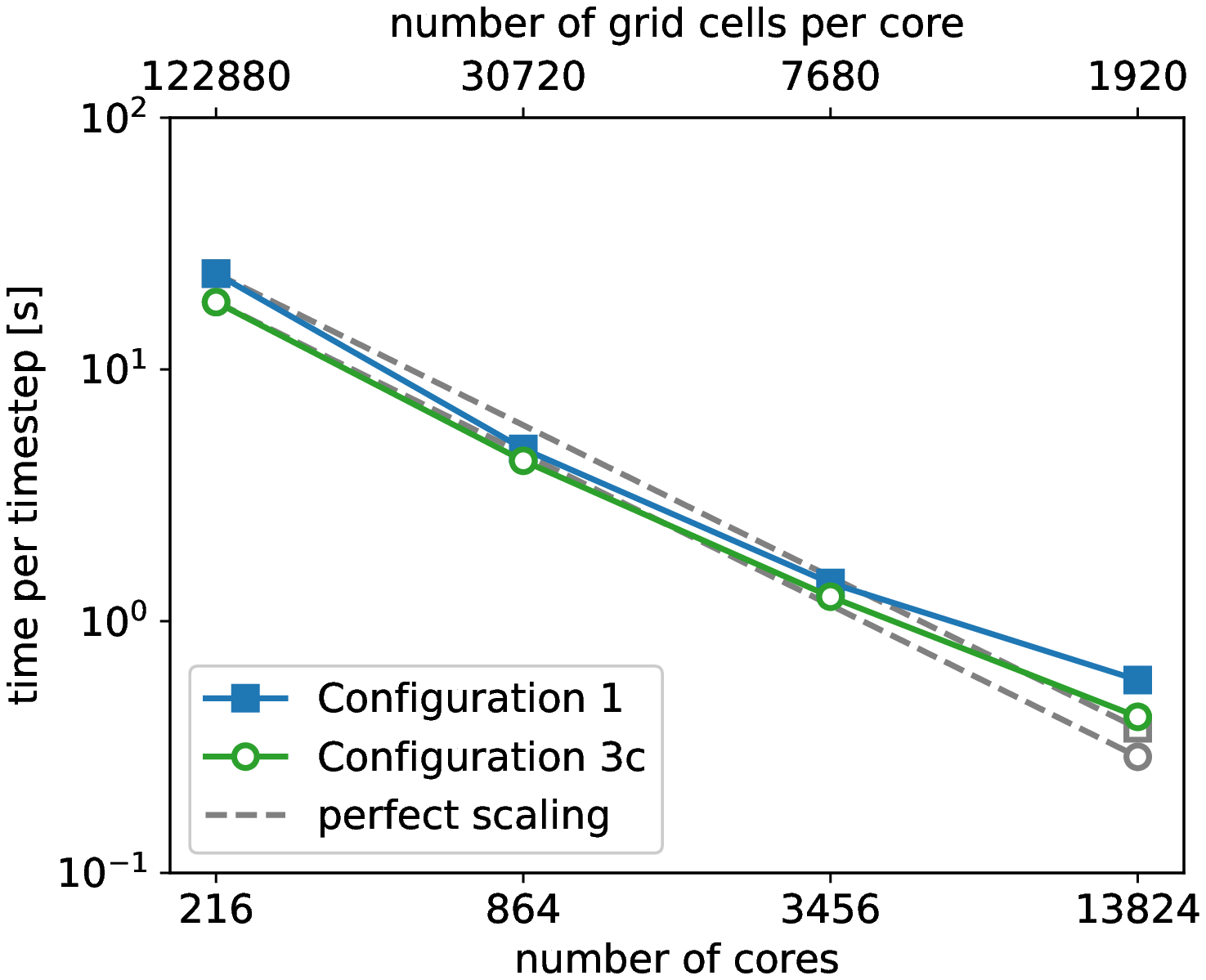}}
  \caption{Time per timestep for the pressure multigrid setup (Configuration 1) and the hybridised multigrid solver (Configuration 3c) under weak scaling (left) and strong scaling (right).}
  \label{fig:scaling-runtime}
\end{figure}

\begin{figure}[!ht]
  \centering
  \subfloat[][]{\includegraphics[width=0.5\textwidth]{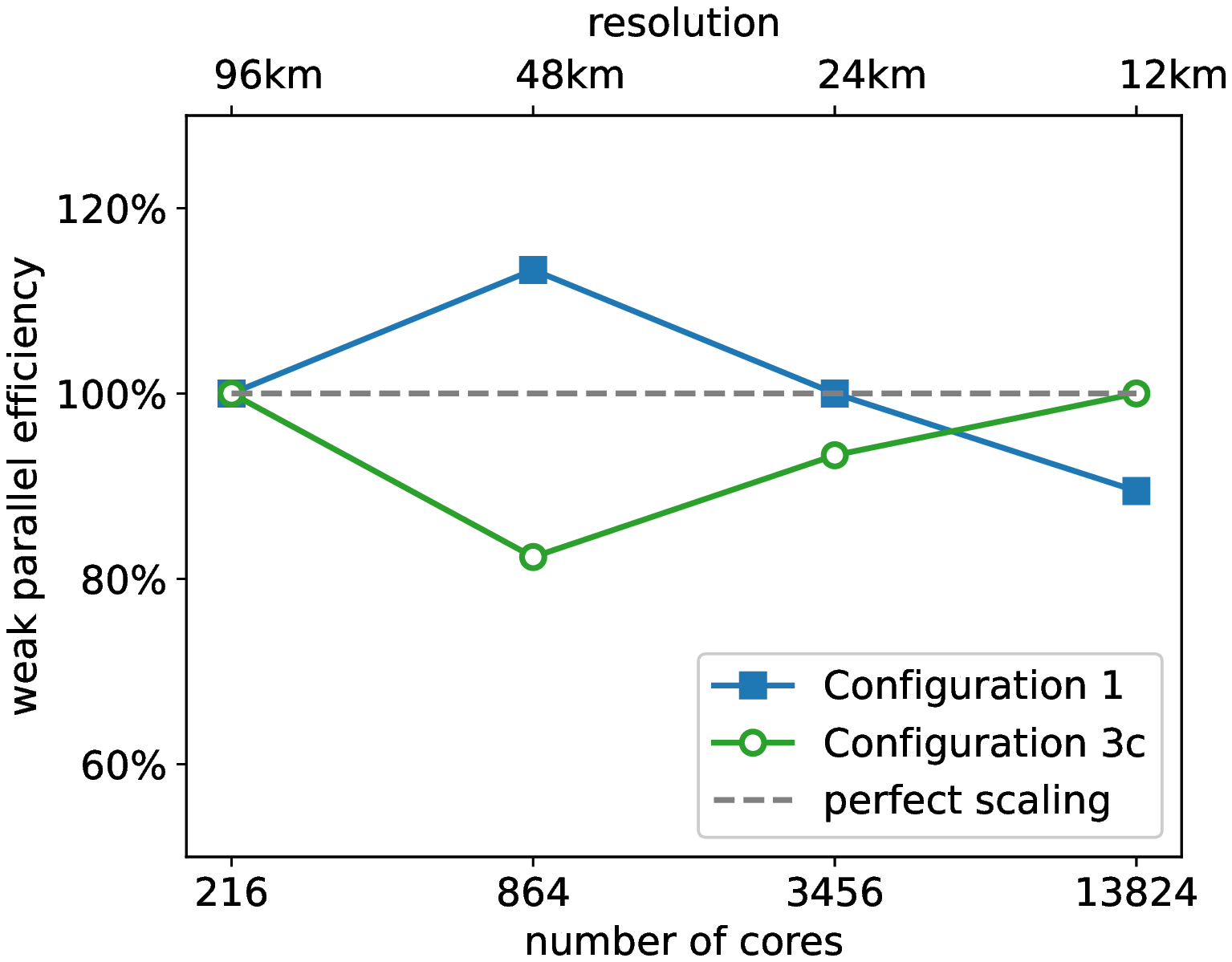}}
  \subfloat[][]{\includegraphics[width=0.5\textwidth]{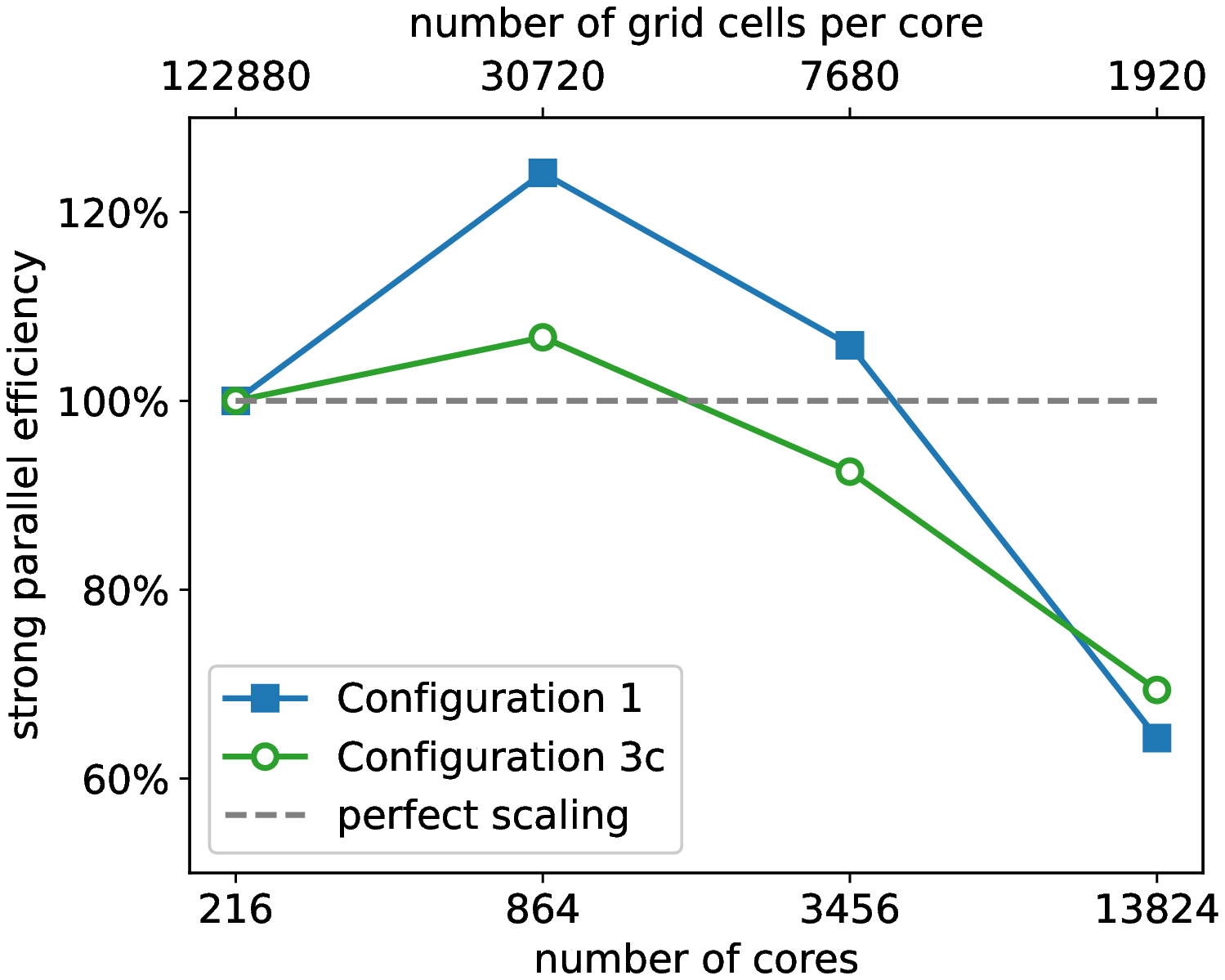}}
  \caption{Weak scaling (left) and strong scaling (right) efficiency as defined in Eq. \eqref{eqn:parallel_efficiency} for the pressure multigrid setup (Configuration 1) and the hybridised multigrid solver (Configuration 3c).}
  \label{fig:scaling}
\end{figure}

\begin{figure}[!ht]
  \centering
  \subfloat[][]{\includegraphics[width=0.5\textwidth]{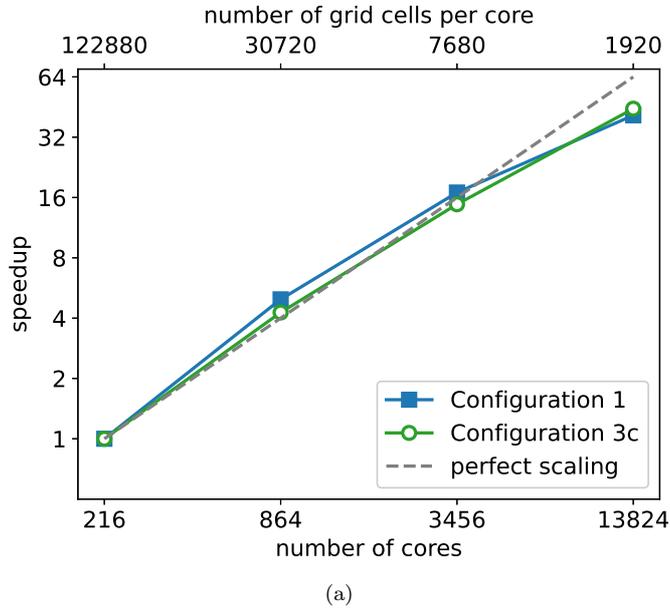}}
  \caption{Parallel speedup of the time per timestep for the pressure multigrid setup (Configuration 1) and the hybridised multigrid solver (Configuration 3c).}
  \label{fig:strong_speedup}
\end{figure}

The strong scaling analysis is carried out for 6, 24, 96 and 384 nodes on a C384 mesh with $6\times 384\times384\times 30=26,542,080$ grid cells and a resolution of 24km, fixing the timestep size to $\Delta t=900\text{s}$. As the number of processors increases, the local volume decreases from $64^2$ columns on 6 nodes to $8^2$ columns on 384 nodes; note that since four multigrid levels are used, this corresponds to a local volume of just 1 column (30 cells) per  MPI region on the coarsest level of the multigrid hierarchy.  More details on the setup for the strong scaling runs can be found in Table \ref{tab:setup_strong_scaling}.
\begin{table}
  \begin{center}
    \begin{tabular}{|r|r|c|c|c|r|r|}
      \hline
      \multicolumn{2}{|c|}{} & \multicolumn{2}{c|}{\# columns per core} & \multicolumn{2}{c|}{\# grid cells per core}                                    \\
      \hline
      \# nodes               & \# cores                                 & finest                                      & coarsest    & finest  & coarsest \\
      \hline
      6                      & 216                                      & $64\times 64$                               & $8\times 8$ & 122,880 & 1,920    \\
      24                     & 864                                      & $32\times 32$                               & $4\times 4$ & 30,720  & 480      \\
      96                     & 3,456                                    & $16\times 16$                               & $2\times 2$ & 7,680   & 120      \\
      384                    & 13,824                                   & $8\times 8$                                 & $1\times 1$ & 1,920   & 30       \\\hline
    \end{tabular}
  \end{center}
  \caption{Setup of strong scaling runs on the C384 mesh. The number of columns and grid cells per core are given both for the finest and for the coarsest mesh in the multigrid hierarchy.}
  \label{tab:setup_strong_scaling}
\end{table}
The performance of the solver under both weak- and strong scaling is shown in Figure \ref{fig:scaling-runtime}, in both cases we report the average time per timestep. Denoting this time by $t_{\text{step}}(p)$ when running on $p$ nodes, the corresponding parallel efficiencies relative to a 6-node run can be defined as
\begin{xalignat}{2}
  \text{weak parallel efficiency} &= \frac{t_{\text{step}}(6)}{t_{\text{step}}(p)},&
  \text{strong parallel efficiency} &= \frac{6\cdot t_{\text{step}}(6)}{p\cdot t_{\text{step}}(p)}.
  \label{eqn:parallel_efficiency}
\end{xalignat}
These efficiencies are shown in Figure \ref{fig:scaling}. For the strong scaling experiment we also plot the speedup $t_{\text{step}}(p)/t_{\text{step}}(6)$ relative to a 6-node run in Figure \ref{fig:strong_speedup}.

At this point it should be stressed that the parallel infrastructure for the LFRic model is under continuous development and optimising parallel performance for the solver is beyond the scope of this paper. At all resolutions the timings of the two considered setups are similar, but the hybrid solver generally exhibits slightly better scaling and results in a shorter runtime at the highest node counts. Interestingly, the strong scaling efficiency for both solver configurations exceeds $100\%$ for 24 nodes but drops for larger node counts, with the pressure multigrid solver exhibiting a more significant drop in efficiency. This effect is more pronounced for the strong scaling experiment, where it can possibly be explained by the fact that improved cache efficiency due to the smaller local volume counteracts a drop in performance due to communication overheads. This does not explain the $>100\%$ efficiency for Configuration 1 on 24 nodes and further work is required to understand this effect. The weak scaling efficiency of the hybrid solver is almost $100\%$ on the largest core count, whereas it drops to around $90\%$ for the pressure multigrid solver (Configuration 1). The strong scaling efficiency is significantly reduced on 384 nodes. Recall that at this point each core will only own the unknowns in a single vertical column with 30 grid cells. On the finest grid, which accounts for a larger fraction of the runtime, the number of unknowns is $8\times 8\times 30 = 1920$. Common experience shows that codes usually stop to scale when the number of local unknowns is at the order of several thousands (see \cite{Fischer2015} and the discussion in \cite[Section 5.2]{Maynard2020}), so our observations are in line with this.

\section{Conclusion}\label{sec:conclusion}
In this work we introduced a new hybridisable finite element discretisation for the solution of the linear equation that arises in semi-implicit timestepping in the LFRic finite element core. A non-nested multigrid method is used to precondition the resulting linear system in trace space. Our numerical experiments show that the new method is competitive with the pressure multigrid method in \cite{Maynard2020}. Both methods show good weak- and strong scaling on parallel architectures.
\subsection{Discussion}
One reason why the approximate Schur complement solver with multigrid preconditioner (Configuration 1) is still competitive with our hybridised non-nested multigrid solver (Configuration 3c) is that, at lowest order, the number of unknowns and the sizes of the local stiffness matrices are comparable. For example, bearing in mind that the velocity unknowns are shared between neighbouring cells, there are four unknowns per cell for the $2\times 2$ system in Eq. \eqref{eqn:2x2system} and the local stiffness matrices are of size $7\times 7$. For the trace system in Eq. \eqref{eqn:hybrid_equation}, there are three unknowns per cell (the trace unknowns on the six facets are shared) and the local stiffness matrices are of size $6\times 6$.

For general polynomial degree $p$, the number of pressure- and velocity unknowns can be worked out using, for example \cite{Arnold2014}. For the mixed system in Eq. \eqref{eqn:2x2system}, the average number of unknowns per cell grows as $\mathcal{O}(p^3)$ and hence the local matrices have $\mathcal{O}(p^6)$ entries. Since the trace unknowns for the hybridised system are associased with the lower-dimensional facets, the number of degrees of freedom per cell grows only as $\mathcal{O}(p^2)$ and the cell-local matrices have $\mathcal{O}(p^4)$ entries\footnote{Construction of the right-hand side in Eq. \eqref{eqn:hybrid-rhs} and recovery of the velocity and pressure with Eq. \eqref{eqn:hybrid-recovery} still requires multiplication with matrices of size $\mathcal{O}(p^6)$, but this is only necessary once per solve.}; explicit numbers for polynomial degrees up to $p=3$ are given in Table \ref{tab:ndofs}.
\begin{table}
  \begin{center}
    \begin{tabular}{|c|rcrcrr|rr|}\hline
      degree & \multicolumn{6}{c|}{number of unknowns per cell} & \multicolumn{2}{c|}{cell-local matrix size}                                                                                          \\
      $p$    & $N_p$                                            & $+$                                         & $N_u$ & $=$ & $N_{\text{mixed}}$ & $N_{\lambda}$ & mixed            & hybridised       \\\hline
      $  0$  & $  1$                                            & $+$                                         & $  3$ & $=$ & $  4$              & $   3$        & $   7\times   7$ & $   6\times   6$ \\
      $  1$  & $  8$                                            & $+$                                         & $ 24$ & $=$ & $ 32$              & $  12$        & $  44\times  44$ & $  24\times  24$ \\
      $  2$  & $ 27$                                            & $+$                                         & $ 81$ & $=$ & $108$              & $  27$        & $ 135\times 135$ & $  54\times  54$ \\
      $  3$  & $ 64$                                            & $+$                                         & $192$ & $=$ & $256$              & $  48$        & $ 304\times 304$ & $  96\times  96$ \\\hline
    \end{tabular}
    \caption{Number of unknowns per cell and local matrix sizes for different polynomial degrees $p$, using $DG_p$ pressure elements and $RT_p$ velocity elements \cite{Arnold2014}. The number of pressure unknowns is denoted with $N_p$, the number of velocity unknowns is $N_u$ and the number of mixed unknowns is $N_{\text{mixed}}=N_p+N_u$; $N_\lambda$ denotes the number of trace unknowns. Shared unknowns associated with facets of the mesh are weighted with a factor $1/2$ in $N_u$ and $N_{\lambda}$.}
    \label{tab:ndofs}
  \end{center}
\end{table}
Assuming that the multigrid solver achieves $p$-independent convergence, we therefore expect the hybridised solver to be more efficient, simply because application of the matrix $S$ in Eq. \eqref{eqn:hybrid_equation} is cheaper than multiplication by the $2\times 2$ matrix in Eq. \eqref{eqn:2x2system}.
\subsection{Future work}
There are several avenues for future research. Firstly, the discussion in the previous section implies that the relative runtime of the three different solver configurations studied here will likely change for higher order discretisations. An increase in discretisation order will also open the door for additional optimisations, such as the use of sum-factorisation techniques in matrix-free implementations (see e.g. \cite{vos2010h,kronbichler2012generic}), which can have a significant effect on solver performance. Secondly, the multigrid algorithm for the second order Dirichlet problem in \cite{cockburn2014multigrid} uses a conforming, piecewise linear P1 coarse level space whereas in this work we chose to use a piecewise constant space for practical reasons. Extending our non-nested multigrid algorithm to a P1 coarse level space could potentially be more efficient, but requires substantial changes to the LFRic infrastructure. This change and a theoretical analysis similar to \cite{cockburn2014multigrid} should be pursued in a subsequent publication. Currently, the pressure operator in the non-nested multigrid algorithm is constructed as an approximate Schur-complement of the original saddle point system. An alternative approach, which was used in \cite{betteridge2021multigrid}, would be to form the Schur-complement of the continuous equations before spatial discretisation and then re-discretise in pressure space. While previous work \cite{mitchell2016high} has shown that the bespoke multigrid algorithms for finite element discretisations of atmospheric model problems are competitive with algebraic multigrid (AMG), it is certainly worth re-visiting this in the context of the hybridisable discretisations considered here. One could, for example, use AMG as a coarse-level solver for Eq. \eqref{eqn:coarse_level_system} in the non-nested two-level method in Algorithm \ref{alg:multigrid} or directly as a solver for the system of Lagrange multipliers in Eq. \eqref{eqn:hybrid_equation}.

Given that time-to-solution is critical in operational forecast models, it is also worth exploring further low-level performance improvements. As discussed in Section \ref{sec:breakdown_times}, a significant fraction of the runtime is spent in tridiagonal solves. When run in parallel, the Thomas algorithm is limited by memory bandwidth, i.e. the time it takes to load the right-hand side and the three matrix diagonals into memory and write the solution vector. It is likely that the matrix columns change very little across the horizontal domain, and one could therefore consider using the same, cached values over small patches of the grid. Although this might lead to a slightly weaker multigrid smoother, this is likely offset by a potentially twofold increase in computational efficiency in the smoother due to reduced memory traffic. A similar optimisation would be to divide all equations by their diagonal, which would then only require loading the off-diagonal matrix entries in the smoother and operator application. However, additional divisions and multiplications by the diagonal will be required before and after intergrid transfer in the multigrid algorithm.

As discussed in Section \ref{sec:parallel_scaling}, future operational configurations will likely require the model to run on $10^5-10^6$ cores. In this paper we have demonstrated that the multigrid solver shows promising scaling to $13824$ cores. Since the model is still under development and computational hardware is continuously evolving, it is hard to predict how the solver will scale to very large core counts. Further work is required to extend the parallel scaling studies and to verify that the model will be competitive at operational global resolutions of around $5\text{km}$.
\section*{Acknowledgements}
This work was funded by two EPSRC Impact Acceleration Awards (grant numbers EP/R511547/1, EP/R51164X/1) and an EPSRC PostDoctoral Award (grant number EP/W522491/1) of Matthew Griffith.
\ifpreprint 
\else 
  \printendnotes
\fi 

\ifpreprint 
  \bibliographystyle{unsrt}
\fi 

\appendix
\section{Implementation in the LFRic Solver API}\label{sec:appendix_code_architecture}
\begin{figure}[H]
  \begin{center}
    \includegraphics[width=0.9\linewidth]{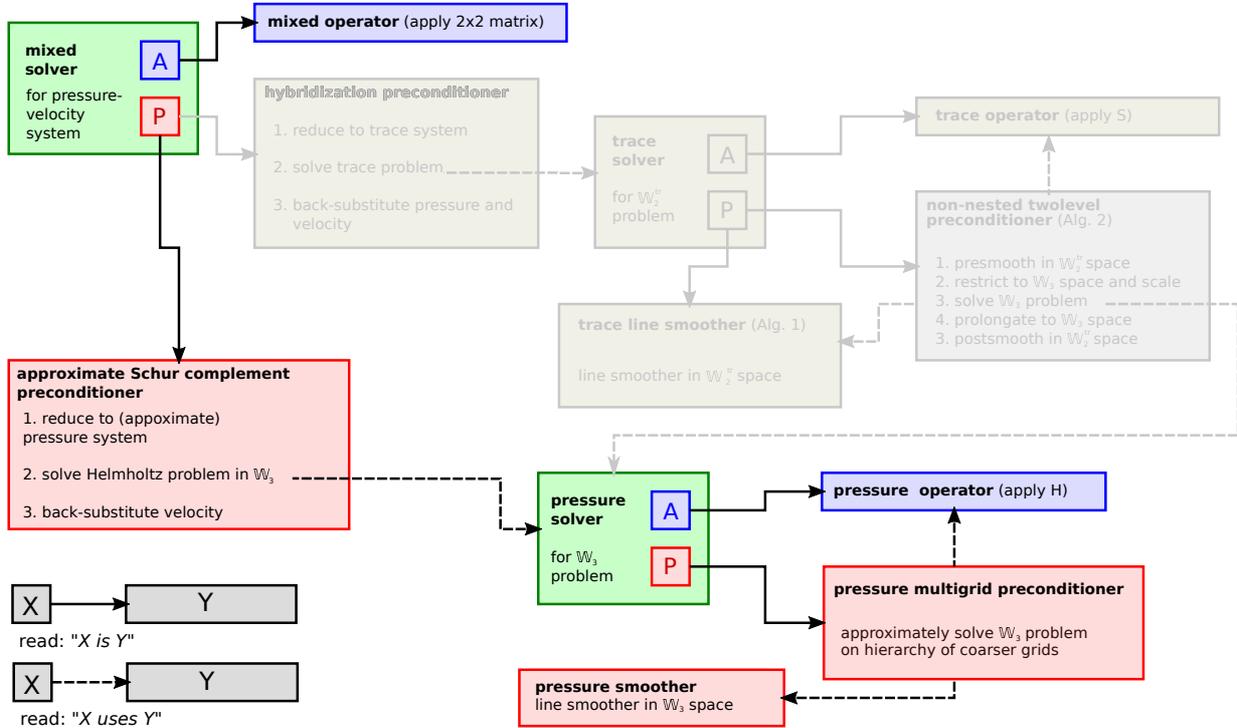}
    \caption{Solver API diagram of solver Configuration 1, the approximate Schur complement solver described in Section \ref{sec:approximate_schur_multigrid}.}
    \label{fig:solver_api_schurcomplement_multigrid}
  \end{center}
\end{figure}

\begin{figure}[H]
  \begin{center}
    \includegraphics[width=0.9\linewidth]{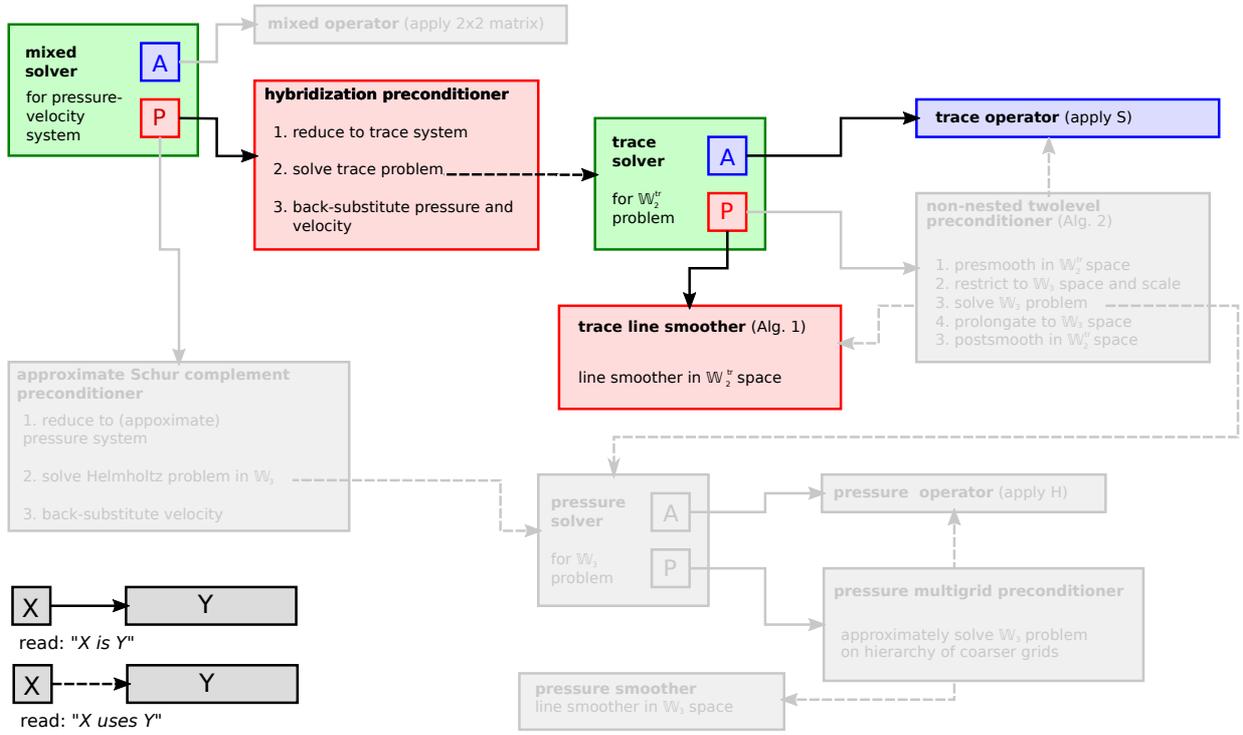}
    \caption{Solver API diagram of solver Configuration 2, the hybridised solver described in Section \ref{sec:hybridisation} and Section \ref{sec:nonnested_multigrid}, using a few iterations of the linesmoother (Algorithm \ref{alg:hybrid_linesmoother}) to solve the trace system in Eq. \eqref{eqn:hybrid-S}.}
    \label{fig:solver_api_hybridized_singlelevel}
  \end{center}
\end{figure}

\begin{figure}[H]
  \begin{center}
    \includegraphics[width=0.9\linewidth]{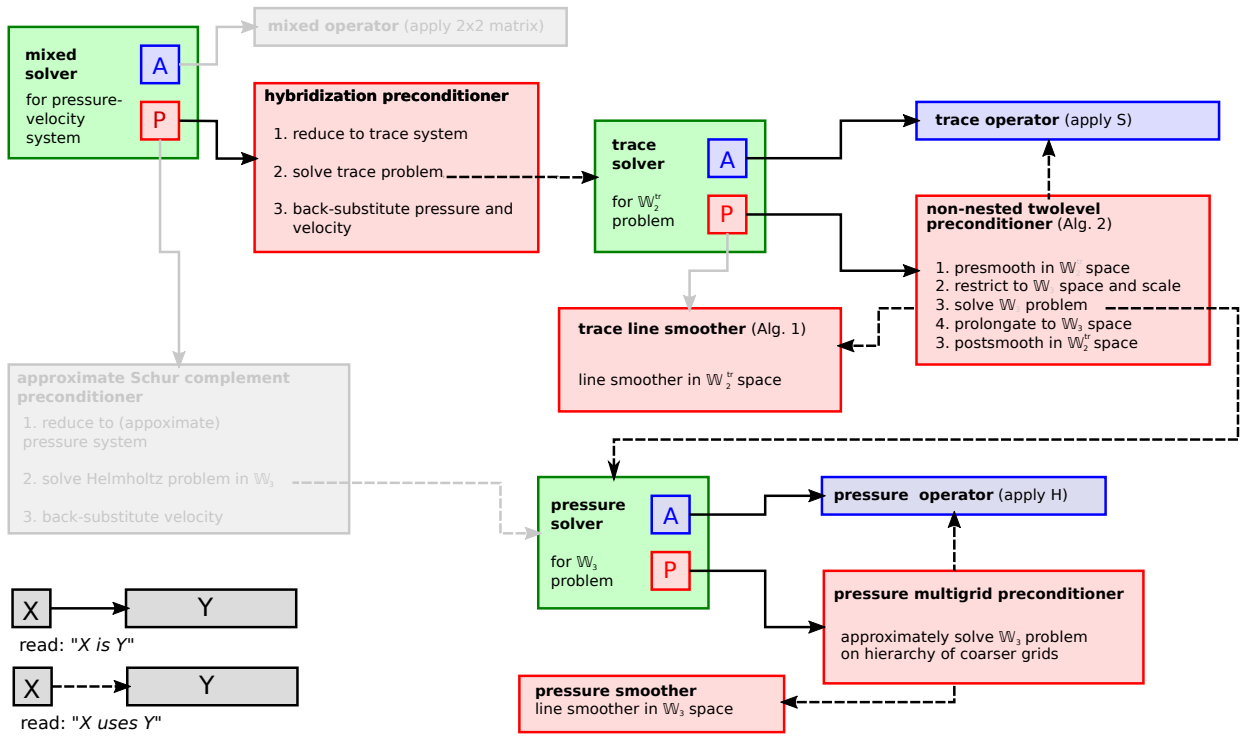}
    \caption{Solver API diagram of solver Configuration 3, the hybridised solver described in Section \ref{sec:hybridisation} and Section \ref{sec:nonnested_multigrid}, using the non-nested multigrid method (Algorithm \ref{alg:multigrid}) to precondition the trace system in Eq. \eqref{eqn:hybrid-S}. The coarse level pressure system is solved with the multigrid method described in Section \ref{sec:approximate_schur_multigrid}.}
    \label{fig:solver_api_hybridized_multigrid}
  \end{center}
\end{figure}

Figures \ref{fig:solver_api_schurcomplement_multigrid}, \ref{fig:solver_api_hybridized_singlelevel} and \ref{fig:solver_api_hybridized_multigrid} show the LFRic Solver API implementation of the three different solver configurations described in Section \ref{sec:solver_configurations}. Each iterative solver instance (shown in green in the diagrams) contains operator objects (blue) and preconditioner objects (red). Since the types for iterative solvers, operators and preconditioners are derived from their respective abstract base types, they conform to a well-defined interface and can be plugged together in the code. In addition to common Krylov subspace methods such as BiCGStab, it is also possible to only apply the preconditioner in a specialised ``pre-only'' iterative solver type. This approach is used in the hybridised solvers in Figures \ref{fig:solver_api_hybridized_singlelevel} and \ref{fig:solver_api_hybridized_multigrid} since the pressure- and velocity unknowns that are reconstructed from the trace field are the exact solutions of the original, non-hybridised system in Eq. \eqref{eqn:2x2system}.

\section{Solver Parameters}\label{sec:solver_parameters}
\begin{figure}[!ht]
  \centering
  \includegraphics[width=0.8\textwidth]{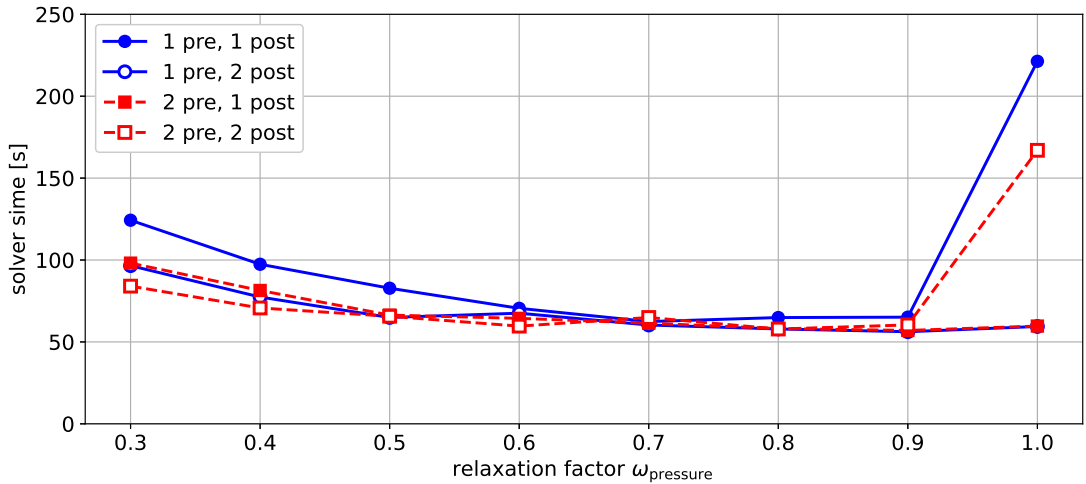}
  \caption{Average time spent in the solver for Configuration 1 (pressure multigrid solver) as a function of the relaxation factor $\omega_{\text{pressure}}$ for different numbers of pre- and post- smoothing steps.}
  \label{fig:pressure_mg_params}
\end{figure}

The performance of the sophisticated solvers considered in this work depends on a number of parameters such as the number of pre- and post- smoothing steps and the relaxation factor $\omega$ in the Jacobi iteration in Algorithm \ref{alg:hybrid_linesmoother}. We refer to the corresponding relaxation factor for the pressure multigrid algorithm in \cite{Maynard2020} as $\omega_{\text{pressure}}$. To verify that the chosen parameters are sensible, we investigated the runtime of the solver in Configurations 1, 2 and 3c for a range of parameters; the code was run for the same single node setup as used in Section \ref{sec:results_single_node}. For Configuration 1 (pressure multigrid) the number of pre- and post- smoothing iterations are set to 1 or 2 and the relaxation factor is varied between $\omega_{\text{pressure}}=0.3$ and $\omega_{\text{pressure}}=1.0$. The time spent in the solver for each setting is shown in Figure \ref{fig:pressure_mg_params}.
\noindent
For $\omega_{\text{pressure}}\leq 0.9$ the measured times show very little variation, indicating that performance depends only weakly on the relaxation factor and the number of pre- and post- smoothing steps. The only exception is the setup consisting of only a single pre- and post- smoothing iteration, which is distinctly slower than the other configurations for smaller values of $\omega_{\text{pressure}}$.

\begin{figure}[!ht]
  \centering
  \includegraphics[width=0.8\textwidth]{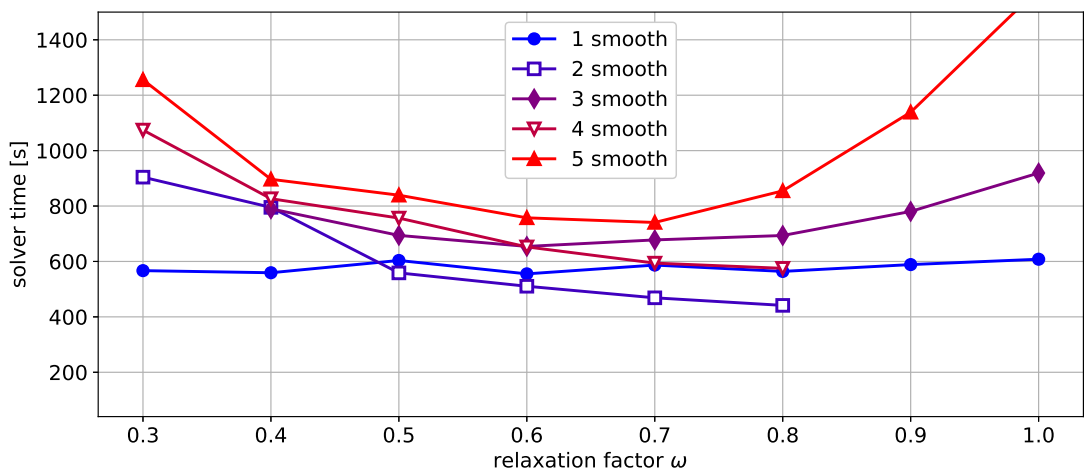}
  \caption{Average time spent in the solver for Configuration 2 (hybridised single-level solver) as a function of the relaxation factor $\omega$ for different numbers of iterations of the Jacobi smoother that is used as a preconditioner. For an even number of iterations the solver only converges for $\omega\le 0.8$; for three smoothing iterations the solver does not converge for $\omega=0.3$.}
  \label{fig:hybrid_ls_params}
\end{figure}

For Configuration 2 (hybridised single-level) the range of relaxation factors $\omega$ which produce good performance is reduced to $0.4\le\omega\le0.8$. There is a tradeoff between the efficiency of the preconditioner, which improves with the number of smoother iterations, and the cost per iteration in the Krylov solver, which favours fewer smoothing steps. Two smoothing iterations is generally the optimal choice. Figure \ref{fig:hybrid_ls_params} shows that the solver cost gradually increases as more iterations are used.

\begin{figure}[!ht]
  \centering
  \includegraphics[width=0.8\textwidth]{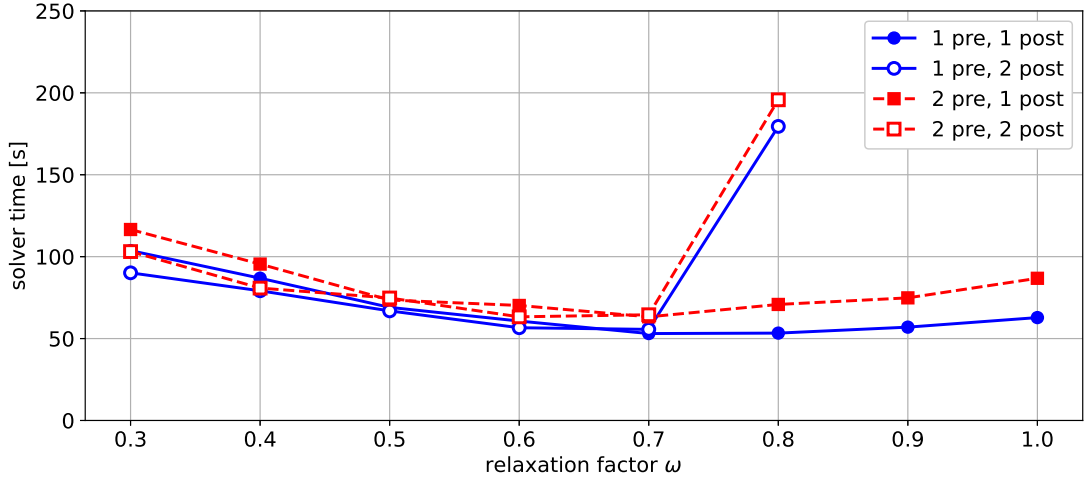}
  \caption{Average time spent in the solver for Configuration 3c (hybridised multigrid solver) as a function of the relaxation factor $\omega$ for different numbers of pre- and post- smoothing steps of the non-nested two-level method. The solver does not converge for relaxation factors larger than $\omega=0.8$ if the number of postsmoothing steps exceeds one.}
  \label{fig:hybrid_mg_params}
\end{figure}

\noindent
Finally, the experiment is repeated for Configuration 3 (hybridised multigrid). In this case the coarse level pressure multigrid preconditioner is fixed to use 1 pre- and 2 post-smooth iterations with $\omega_{\text{pressure}}=0.9$ and the number of smoothing iterations and  the relaxation factor in the nonested two-level preconditioner (Algorithm \ref{alg:multigrid}) is varied. Figure \ref{fig:hybrid_mg_params} shows the time spent in the solver as a function of the relaxation factor and for up to three pre-smoothing steps. As long as $\omega \leq 0.7$ there is little difference in cost between the configurations, with a slight improvement in performance as $\omega$ increases. Unless the number of post-smooth iterations is limited to one, performance drops substantially for $\omega=0.8$ and the solver fails to converge for larger relaxation factors. The optimal setup is generally to use 1 pre- and 2 post smooth iterations.

\end{document}